\documentclass[npg,ms]{copernicus}
\usepackage{times}
\usepackage{amsmath}
\usepackage{subfigure}

\printfigures            
\newcommand{\bd}{\begin{displaymath}}
\newcommand{\ed}{\end{displaymath}}
\newcommand{\be}{\begin{equation}}
\newcommand{\ee}{\end{equation}}
\newcommand{\ba}{\begin{array}}
\newcommand{\ea}{\end{array}}
\newcommand{\bal}{\begin{align}}
\newcommand{\eal}{\end{align}}
\newcommand{\bpm}{\begin{pmatrix}}
\newcommand{\epm}{\end{pmatrix}}

\def \O {\mathcal{O}}
\def \tp {\tilde{\psi}}
\begin{document}

\title{ Continuing dynamic assimilation of the inner region data in 
hydrodynamics modelling: Optimization approach}

\author[1]{F. I. Pisnichenko}
\author[2]{I. A. Pisnichenko}
\author[1]{J. M. Martinez}
\author[1]{S. A. Santos}

\affil[1]{Universidade Federal de Campinas, Campinas, SP, Brazil}
\affil[2]{Centro de Previs\~{a}o de Tempo e Estudos Clim\'{a}ticos/Instituto 
         Nacional de Pesquisas Espaciais, Cachoeira Paulista, SP, Brazil}


\runningtitle{Continuing dynamic assimilation of the inner region data}

\runningauthor{F. I. Pisnichenko et al.}

\correspondence{F. I.Pisnichenko \\ (feodor@ime.unicamp.br)}

\received{}
\pubdiscuss{} 
\revised{}
\accepted{}
\published{}


\firstpage{1}

\maketitle

\begin{abstract}
In meteorological and oceanological studies the classical approach for finding 
the numerical solution of the regional model consists in formulating and 
solving the Cauchy-Dirichlet problem. The related boundary conditions are 
obtained by linear interpolation of data available on a coarse grid (global 
data), to the boundary of regional model. Errors, in boundary conditions, 
appearing owing to linear interpolation may lead to increasing errors in 
numerical solution during integration. The methods developed to reduce these 
errors deal with continuous dynamic assimilation of known global data available 
inside the regional domain. Essentially, this assimilation procedure performs 
a nudging of large-scale component of regional model solution to large-scale 
global data component by introducing the relaxation forcing terms  into 
the regional model equations. As a result, the obtained solution is not a valid 
numerical solution of the original regional model. In this work we propose 
the optimization approach which is free from the above-mentioned shortcoming. 
The formulation of the joint problem  of finding the regional  model solution 
and data assimilation, as a PDE-constrained optimization problem, gives 
the possibility to obtain the exact numerical solution of the regional model. 
Three simple model examples (ODE Burgers equation, Rossby-Oboukhov equation, 
Korteweg-de Vries equation) were considered in this paper. The result of 
performed numerical experiments indicates that the optimization approach can 
significantly improve the precision of the sought numerical solution, even in 
the cases in which the solution of Cauchy-Dirichlet problem is very sensitive 
to the errors in the boundary condition. 
\end{abstract}

\introduction
Studying and modelling different physical processes frequently require to solve 
the Cauchy-Dirichlet problem. In geophysical investigations, for solving of 
the Cauchy-Dirichlet problem, numerical methods are usually applied.
The discrete form of equations and initial and boundary values at 
the points of the grid are traditionally used in these methods.
The problem is generally solved by using a proper numerical scheme. 
However, both initial and boundary  values, which are obtained
from measurements or other model outputs, contain errors, which can reach 
$30 \%$ from the true values.  

For some problems, the solution can be very sensitive to these errors. 
At the same time some values  of the sought solution at some points 
of the inner region, along with initial and boundary data, are often  available. 
These additional data also have errors.  The question arises: Is it possible 
to improve the accuracy of the solution of the Cauchy-Dirichlet problem 
by using these additional data (which may contain errors)?
For some well-known equations of mathematical physics and dynamic meteorology
we will show here that the use of additional information on the solution values
within the integration region can noticeably improve the accuracy of the sought 
solution.

An improvement of  the solution accuracy of the Cauchy-Dirichlet
problem is extremely important in meteorology because  this is the problem 
of regional weather forecast. A distinction between two  types of atmospheric 
forecast models needs to be done. These are global models, making forecast 
for the whole earth, and regional models, which produce a forecast for a limited 
region. One of  basic differences between these model types is the grid 
resolution. Global models use a low resolution space grid and regional models 
operate on a more dense mesh. The reason for the existence of these two model types is, 
basically, computational. Even nowadays it is not possible in acceptable time period
to integrate global models with the detailed physics and  with the space 
resolution of regional models. It is worth noticing that global and regional 
models include different physical processes. Global models describe large-scale 
slow-time varying  processes, with the time period of more than 3  hours and 
the space scale larger than 60 km. Regional models can simulate the evolution of 
mesometeorological  fast processes (small cyclones, storms, tornados), with 
the time period of less than 3 hours and the spatial extent  between 2 and 60 km. 

These models are described mathematically by systems of nonlinear partial
differential equations. To solve the system corresponding to a global model 
we need only the initial condition and bottom and upper  boundary conditions, 
as the domain is the whole sphere. To get the solution of a system corresponding to
a regional model we also need lateral boundary conditions. Generally, the lateral
boundary conditions for a regional model are obtained from the global model. 
Global model data corresponding  to the domain of a regional model lateral boundary
are interpolated into a regional mesh and then are used for finding of the 
Cauchy-Dirichlet problem solution.

We should notice that the lateral boundary conditions obtained from the global model do not 
provide any information on  the structures of the scales smaller than the size of 
the mesh of the global model. In fact, global models
do not distinguish a local meteorological phenomenon of characteristic length
scale smaller than $30$ km, because the space step is greater than this value.
On the other hand, the parts of  space and time spectrums of a regional model 
solution which correspond to large-scale structures and long-period processes are 
in worse accordance with observations than those of a global model. This is related 
to the fact that a regional model does not have the information about the phenomena
that occur outside its domain and therefore cannot describe with necessary 
precision (because the lateral boundary data are in low resolution discrete space 
and time grid) the impact of these phenomena on the evolution of processes inside 
the integration domain.

For example, the weakly regular, low frequency  oscillations of the sea surface
temperature near the Peru coast, known as El Ni\~no and La Ni\~na events, have a large
influence on the climate of Brazil. The circulation regimes over the Brazilian 
territory noticeably  differ during these events.  As a result, the precipitation
rate in the South of Brazil during the La Ni\~na periods is two times smaller than during 
the El Ni\~no periods. Regional models driven, for example, by the Reanalysis data 
do not reproduce with confidence this distinction between the El Ni\~no and La Ni\~na 
regimes. The reason of this inconsistency is, probably, poor space and time 
data resolution on the regional model boundaries. Mathematically, it means that 
we are solving boundary value problems with significant errors on the boundaries 
and the solution is sensitive to these errors.

The spectral nudging technique is one of methods proposed to use additional data from 
the inner domain in order to reduce boundary error influence and to improve 
a solution of initial-boundary value problem  \citep{wald96, storch00, kana05}. 
This method supposes incorporation of the largest internal modes of some meteorological 
variables from observations or from a driving model into the regional model solution.  
However the use of the spectral nudging technique requires inserting additional forcing 
terms into the evolution equations of a regional model. Hence the original model 
is corrupted  and its new solution may not be close to the exact sought solution.

In this work we propose a method which is free from the above mentioned 
shortcoming. Firstly, in Section 2 we give the formulation of the Cauchy-Dirichlet 
problem for a regional model and the general formulation of an optimization problem, 
and we show how the latter (with the accessibility of some additional
conditions) can be applied  for finding the solution of the former.
In Section 3,  for some simple equations, we show how the optimization 
approach can be used to obtain a solution of the initial boundary 
value problem. The equations considered here include 
the ordinary differential equation of Burgers, the one-dimensional, linearized 
Rossby-Oboukhov partial differential equation, and the partial differential 
equation of Korteweg - de Vries. We also compared the sensitivity of the solutions, 
obtained by the traditional and the new approach with respect to the errors 
in the boundary conditions. In Section 4 we discuss the obtained results. 
Appendix A provides more detailed description of the optimization procedure.

\section{Numerical solution of regional problems}

\subsection{The numerical forecast model}

We formulate here the Cauchy-Dirichlet problem as it is formulated for the
regional weather forecast. To find a unique solution for the regional  model 
equations we need the lateral boundary condition for entire time interval on which 
the solution is seeking. We obtain these data from the solution of an outer model which, 
as a rule, is a global model.  We don't know an exact solution of the outer model
(if we did, our problem already would have been solved), but 
we can find its approximate solution in the discrete form:
\begin{equation}\label{globmodel_discr}
\begin{array}{l}
\Delta_t \{\Psi\} = F_d \textrm{,} \\
\Psi(x,0)=Y_{global}(x)\textrm{,} \\
\Psi(x,t)\mid_{x \in B} = \Psi_b(x,t).
\end{array}
\end{equation}
Here $\Delta_t$ is the evolution operator of the discrete model, $\Psi$ is
the vector of  the prognostic functions, $F_d$ are discrete external
forces, $Y_{global}$ is the initial condition, $B$ are upper, lower (and maybe
lateral) boundaries of the global model, and $\Psi_b$ is the boundary condition
for the global model. Let $\Psi_{sl}$  will be the solution of 
(\ref{globmodel_discr}). We suppose that this solution is sufficiently close 
to the solution of hypothetical ideal model which exactly describes the real 
atmospheric processes.
 
The regional model is located in a closed area with boundary $S$ and its
discrete representation can be written in the same manner as for
the global model
\begin{equation}\label{regmodel_discr}
\begin{array}{l}
\delta_t \{G\} = F_{rd} \textrm{,} \\ 
G(x,0)= Y_{local}(x) \textrm{, } \\
G(x,t)\mid_{x \in S} = G_s(x,t) \textrm{,}
\end{array}
\end{equation}
where  $\delta_t$ is the evolution operator of the discrete regional model,
$G$ is the vector of the prognostic functions and $F_{rd}$ are discrete
external forces, $Y_{local}$ is the initial condition, and $G_s$ is 
the boundary condition for the regional model.

Now, we can solve numerically the initial boundary value problem 
(\ref{regmodel_discr}) using the solution $\Psi_{sl}$ of the global model (\ref{globmodel_discr})
for getting  the boundary conditions. The traditional approach to seek  
the solution of (\ref{regmodel_discr}) consists in the interpolation of required data 
from $\Psi_{sl}$  on the regional grid for forming the initial and boundary conditions 
and then applying any numerical method  to solve the model equations. But, as we have 
mentioned above, the boundary conditions contain errors which can strongly corrupt 
the solution. 

The use of the information obtained from the solution of the global 
model equations $\Psi_{sl}$ on the values inside the area of the regional 
model integration for all available time moments can help to overcome this difficulty.

\subsection{Formulation of the optimization problem}

We  assume that the solution of regional model (\ref{regmodel_discr}) cannot strongly differ 
from global model solution $\Psi_{sl}$. Our objective is to find $G_{sl}$ on the fine (regional) mesh 
that satisfies the initial values and the equations of the regional model 
(\ref{regmodel_discr}). The solution $G_{sl}$ also have to be as close as possible 
to the $\Psi_{sl}$ inside the regional model domain.  We can 
formulate this aim as an optimization problem in the following manner
\begin{equation}\label{probger_min}
\begin{array}{rl}
\textrm{Minimize } & d(G,\Psi_{sl})\\
\textrm{subject to \, (s.t.) } & \delta_t \{G\} - F_{rd} = 0,\\
&G(x,0) - Y_{local}(x) = 0.
\end{array}
\end{equation}
Here $d(:,:)$ is the objective function, which represents the distance between 
$G$ and  $\Psi_{sl}$ vectors. In other words, we want to minimize the distance between 
the regional and the global model solutions under conditions that the regional model 
equations and the initial condition are  satisfied. 
Note here that, when we talk about the solution  $G$, we bear in mind the vector 
$G=G(x,t)$  in a grid space of ($x$,$t$) - coordinates, where $t$ corresponds to 
discrete time points of the regional model integration and $x$ to discrete mesh 
points in the regional model area.

There are different approaches to solve the optimization problem 
(\ref{probger_min}). For example, we can apply it to each time level, 
as it is done in finite-differences methods: For the given initial conditions
($t_0=0$), the solution is obtained at $t_1$. Then, considering the solution at 
$t_1$ as the initial condition, the solution is found at  $t_2$ and so on.
On the other hand, we can consider the domain of definition of the problem on 
the mesh including  all time levels. In this case all available information from 
the global model for  the period of integration will be taken  into 
consideration simultaneously. We use here this approach.

\section{ Examples of application of the optimization method to the problems of
small dimension }

\subsection{Burgers' equation}

\begin{figure}
\begin{center}
\includegraphics[width=5cm]{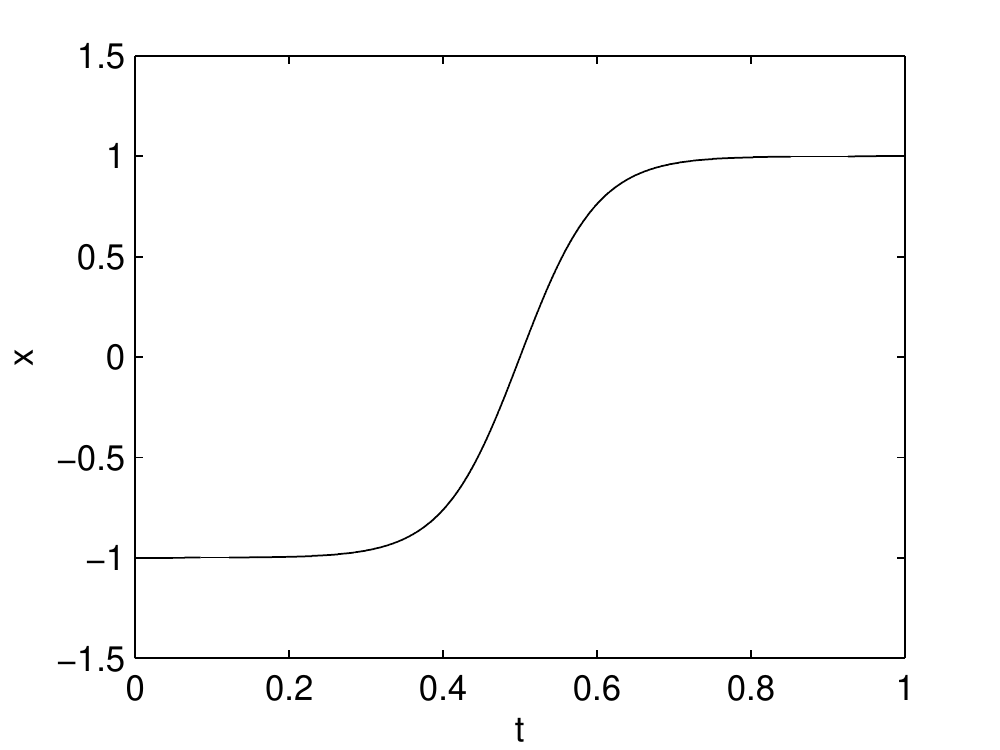}
\caption{\label{fig:fig1}
Analytical solution of the Burgers' equation for $\epsilon = 0.05$
and  the boundary condition:  \\  $x(0)=-1$, $x(1)=1$.}
\end{center}
\end{figure}

We shall demonstrate an application of the optimization method to problems of
small dimension. As a first example we consider the supersensitive boundary value
problem  for the Burgers' equation \citep{bohe96}:
\begin{equation}\label{burgers}
\begin{array}{l}
\epsilon x'' = -xx' \textrm{,}\\
x(0)=-1 \textrm{, } x(T)=1.
\end{array}
\end{equation}
To get the analytical solution of this equation it is enough to integrate 
the left and right sides of the equation (\ref{burgers}) over $t$: 
\bd
\epsilon x' = -\frac{x^2+c}{2}.
\ed
Then, after rewriting the foregoing formula as 
\bd
\epsilon\frac{dx}{x^2+c}=-\frac{dt}{2}
\ed
and integrating the left side over $x$ and the right one over $t$, 
we can write the Burgers' equation solution as
\bd
x =
\begin{cases}
\frac{\displaystyle \sqrt{-c}\left(1+ \exp{\frac{(t+C_2)
\sqrt{-c}}{ \epsilon}}\right)}
{\displaystyle 1-\exp{\frac{ (t+C_2)\sqrt{-c}}{ \epsilon}}},  \, \, \textrm{ for } c
\le 0\textrm{,}\\
\\
\sqrt{c}\tan\left(\frac{\displaystyle (t+C_2)\sqrt{c}}{\displaystyle 2 \epsilon}\right), \, \,   
\textrm{ for } c >0.
\end{cases}
\ed
The graph of the analytical solution for equation (\ref{burgers}) 
with the boundary condition $x(0)=-1$, $x(1)=1$
and $\epsilon = 0.05$ is presented in Figure \ref{fig:fig1}. Numerical solution of
this problem applying, for example, the shooting method coincides 
with great accuracy with the analytical one.

[Insert figure \ref{fig:fig1} here.]

\begin{figure}
\begin{center}
\includegraphics[width=5cm]{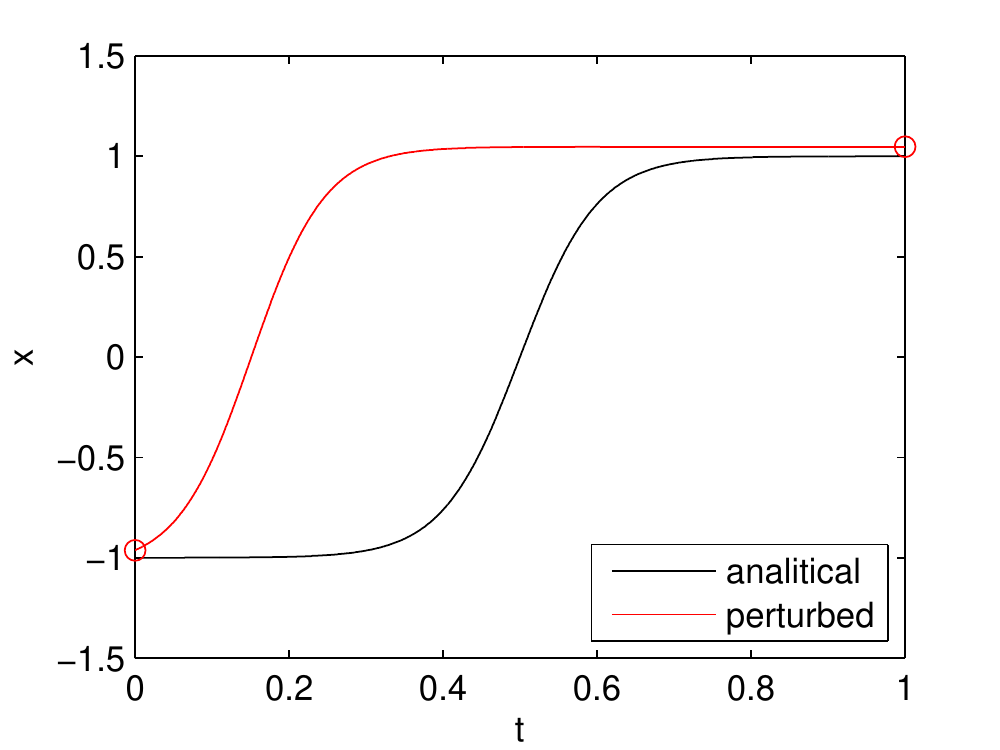}
\caption{\label{fig:fig2}
Forward problem solution with 5\% perturbed  boundary condition of Burgers' equation
in comparison with exact analytical solution. }
\end{center}
\end{figure}

Let us choose several points from  the analytical solution 
and slightly perturb them (till 5\% from its real value).  
This procedure models the boundary  and inner domain data containing errors. 
Using these perturbed data  for the boundary points $x(0)$, $x(1)$  we solve 
the boundary problem (\ref{burgers}) applying  the shooting method. For the step
$ \Delta t=0.01$ the numerical solution is presented in Figure \ref{fig:fig2} 
by red line.
  
[Insert figure \ref{fig:fig2} here.]

As one can see, small perturbations of the boundary condition result in  large 
errors in the solution. Now, let us solve  the optimization problem (\ref{probger_min}) 
using additional information about the  perturbed solution on the points inside 
the domain. Using the same discretization as when solving (\ref{burgers}) by the
traditional shooting method, with  the same $ \Delta t=0.01$ we find the solutions 
which are exhibited in Figure3. 

\begin{figure}
\begin{center}
\includegraphics[width=5cm]{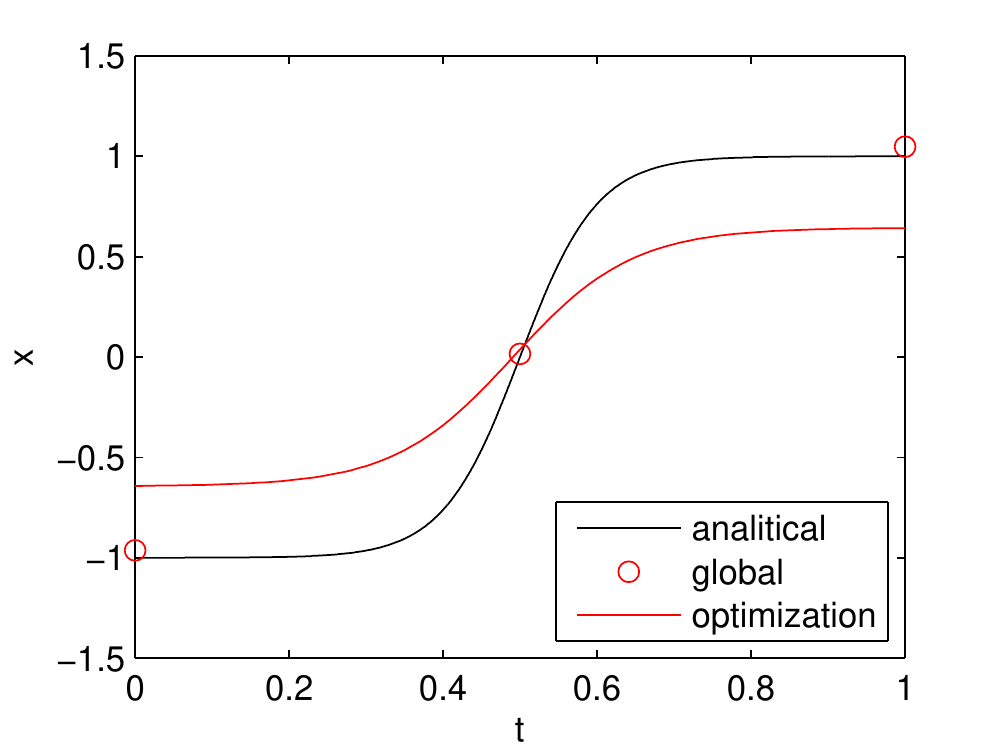}
\includegraphics[width=5cm]{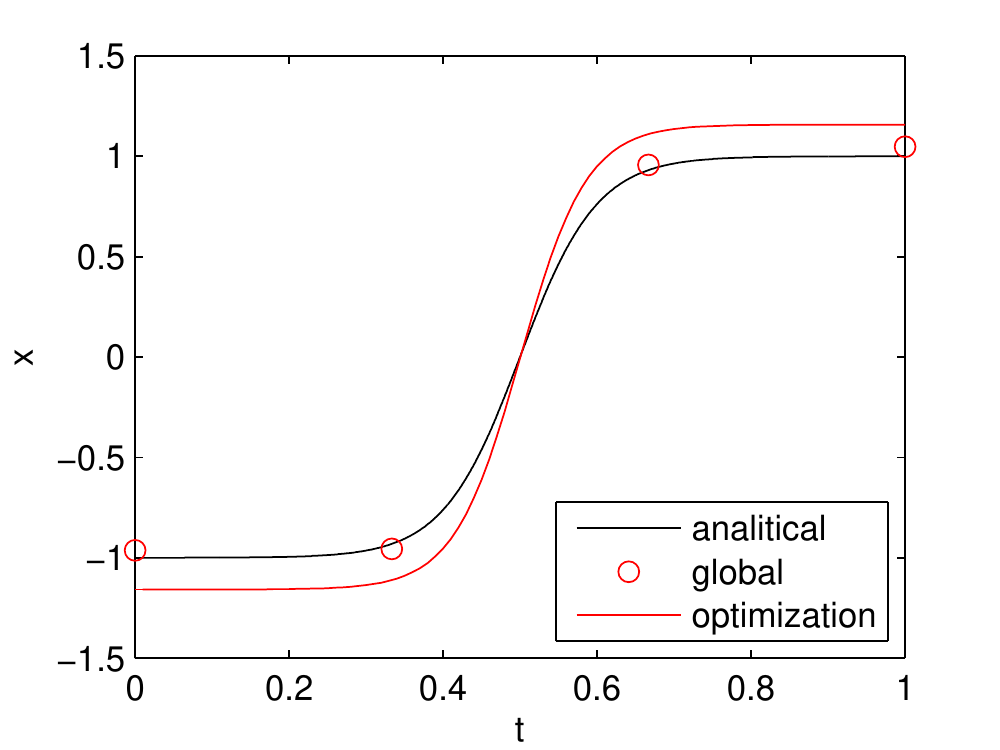}
\includegraphics[width=5cm]{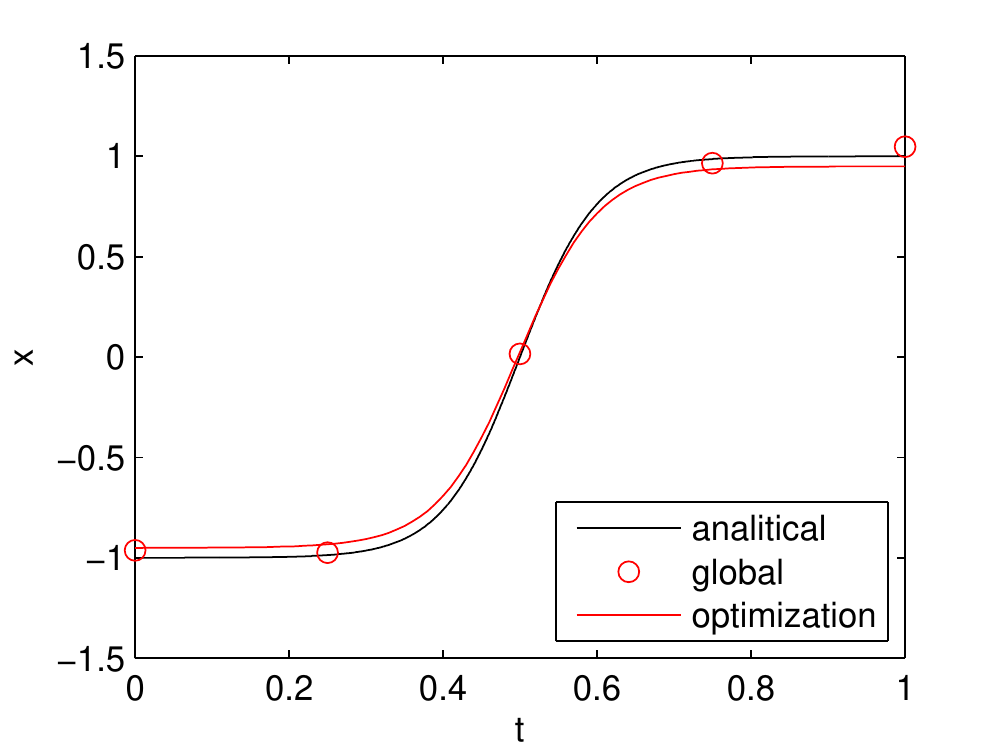}
\caption{\label{fig:fig3}
Optimization problem solution of Burgers' equation for three cases: 3 reference 
points; 4 reference points; and  5 reference points.}
\end{center}
\end{figure}

Figure \ref{fig:fig3} shows the solutions of the optimization problem  for the three 
cases when we used 3, 4 and 5 points of the perturbed solution, respectively.
One can see that all these solutions correspond better to the analytical solution, 
than the solution obtained by the traditional approach. In the case of 
using the three additional inner points, the optimization solution and  the analytical 
solution nearly coincide. This example shows that there are dynamical systems in
which small perturbations ($\le$ 5\%) in the boundary conditions can lead to 
great errors in the final solution. However, if some additional information   
exists, it can be used to improve significantly the sought solution,
applying to the considered problem the optimization theory methods.

[Insert figure \ref{fig:fig3} here.]

\subsection{Rossby-Oboukhov one-dimensional equation}

Here we consider the one-dimensional linear partial differential equation
obtained during the linearization procedure of  the Rossby-Oboukhov 
equation \citep{obouk49}:
\begin{equation}\label{eq_rossby}
\frac{\partial }{\partial t} \left(\frac{\partial^2}{\partial x^2}
- \frac{1}{l_0^2} \right) \psi + \beta \frac{\partial \psi}{\partial x} +
U \frac{\partial^3 \psi}{\partial x^3}=0 \textrm{.}
\end{equation}
Here $\psi$ is the stream function, $f_0 = 10^{-4}$ s$^{-1}$ is mean value of Coriolis 
parameter, $\beta= d f/ d y = 1.6 \cdot 10^{-11}$ s$^{-1}$m$^{-1}$ is mean value of 
meridional gradient of  Coriolis parameter, $l_0=c_0/f_0 = 3 \cdot 10^6$  m is the 
Oboukhov scale, $c_0$ is the sound velocity, and $U$ is the zonal wind, which is 
variated between $0$ and $30$ m/s.

For the periodic boundary condition 
\begin{equation}\label{per_cond}
\psi(0,t)=\psi(L,t),
\end{equation}
where $L$ is the size of integration area (we shall use $L=3 \cdot 10^7$ m), 
the solution of equation (\ref{eq_rossby})  can be written as 
\begin{equation}\label{rossby_sol}
\psi(x,t)=  \sum_{n=1}^N A_n \sin[k_n(x-c_nt)+\phi_n]\textrm{,}
\end{equation}
where 
\begin{equation}\label{rossby_uni_ccoef}
   c_n=U - \frac{\beta +U/l_0^2}{k_n^2 +1/l_0^2}\textrm{,}
\end{equation}   
\bd
   k_n=\frac{2 \pi n}{L},
\ed
and $A_n$ and $\phi_n$ are defined by the initial condition \citep{rossby39}.

For finding the numerical solution it is convenient to rewrite the equation in 
nondimensional form. We choose the following scales $S=6 \cdot 10^6$ m, 
$T=S/V=6 \cdot 10^5$ $s$, $V= 10$ $m/s$. The dependent and independent 
nondimensional variables are defined as follows:
\begin{equation}
\tilde{x}=\frac{x}{S}, \qquad \tilde{t}=\frac{t}{T}, \qquad \tilde{\psi}=\frac{T}{S^2}\psi.
\end{equation},
and equation (\ref{eq_rossby}) may be written in nondimensional form as
\begin{equation}\label{rossby1_ad}
\frac{\partial }{\partial \tilde{t}} \left( \frac{\partial^2}{\partial \tilde{x}^2}
-\frac{1}{b^2}\right) \tilde{\psi}+\beta_0 \frac{\partial \tilde{\psi}}
{\partial \tilde{x}} + U_0 \frac{\partial^3 \tilde{\psi}}{\partial \tilde{x}^3}=0 
\textrm{,}
\end{equation}
where $\frac{1}{b}=\frac{S}{l_0}=2$, $\beta_0=\frac{\beta S^2}{V}=57.6$ and
$U_0=\frac{U}{V} \in [0,3]$. 

For finite-difference discretization we use unconditionally stable scheme with truncation error
$\O(\Delta x^2,\Delta t^2)$ given by
\begin{equation*}
\begin{split}
&\frac{1}{\Delta t} \left(\frac{\tp_{i+1}^{k+1}-2\tp_{i}^{k+1}+\tp_{i-1}^{k+1}}{\Delta x^2} -
\frac{\tp_{i+1}^{k}-2\tp_{i}^{k}+\tp_{i-1}^{k}}{\Delta x^2} -\frac{1}{b^2}
\left(\tp_i^{k+1}-\tp_i^{k}\right)\right)\\
&+\frac{\beta_0}{2}\left(\frac{\tp_{i+1}^{k+1}-\tp_{i-1}^{k+1}}{2\Delta x}
+\frac{\tp_{i+1}^{k}-\tp_{i-1}^{k}}{2\Delta x}\right)\\
&+\frac{U_0}{2}\left(\frac{\tp_{i+2}^{k+1}-2\tp_{i+1}^{k+1}+2\tp_{i-1}^{k+1}-\tp_{i-2}^{k+1}}{2\Delta x^3}
+\frac{\tp_{i+2}^{k}-2\tp_{i+1}^{k}+2\tp_{i-1}^{k}-\tp_{i-2}^{k}}{2\Delta x^3}\right)=0
\end{split}
\end{equation*}

\begin{figure}
\begin{center}
\includegraphics[width=14cm]{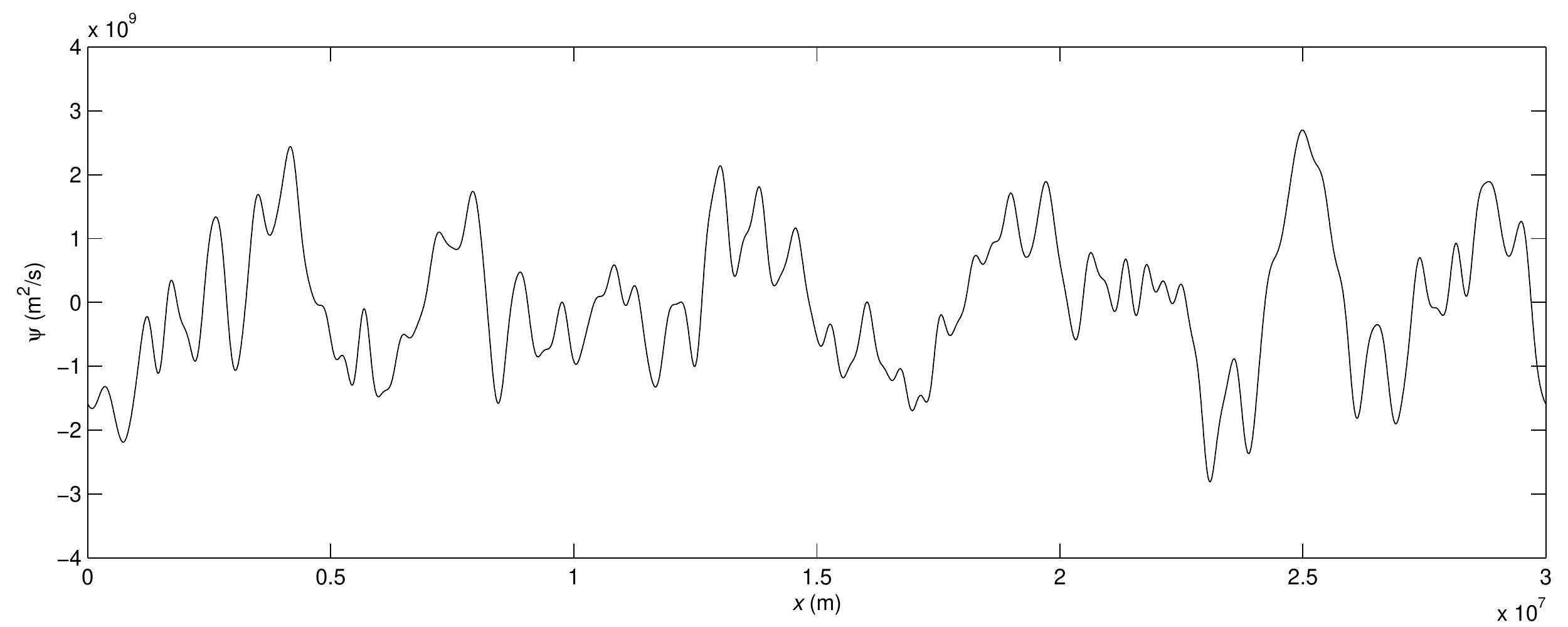}
\caption{\label{fig:fig4}
Analytical solution of Rossby-Oboukhov equation on periodic domain $[0,L]$ at $t=0$ h with
85 modes.}
\end{center}
\end{figure}

At first, we generate a specific analytical solution (\ref{rossby_sol}) containing  
85 modes. Figure \ref{fig:fig4} shows this solution for $t=0$ h.

[Insert figure \ref{fig:fig4} here.]

Note that in Figures we use dimensional values, but in the numerical computations  all 
variables are nondimensional.
 
As a local model we consider the equation (\ref{rossby1_ad}), with initial 
and boundary conditions, defined over $[a,b] \varsubsetneq [0,L]$  (closed interval 
smaller than entire domain). For initial and boundary conditions we will use our specific 
form (see Fig. \ref{fig:fig4}) of \textit{the global} model solution (\ref{rossby_sol}). 
For simulating really encountered problems of atmospheric modelling  we took 
the analytical solution in the points of the the coarse  grid with space step  
$\Delta x = 200$ km and time step $\Delta t = 2$ hour, and perturb its values randomly 
in such a way that a perturbation can reach till 30\% of  its exact  value.

Primarily we find the solution of the forward Cauchy-Dirichlet problem for our local 
model. As local domain we take the interval $[1.8\cdot 10^7,2.4 \cdot 10^7] m$ 
(6000 kilometers) inside the global model interval $[0,3 \cdot 10^7] m$. For the first 
reference experiment the initial and boundary conditions were taken from the exact 
analytical solution (\ref{rossby_sol}). Requiring that the numerical solution with 
the exact initial and boundary condition have to nearly coincide with the analytical 
solution, we took as the maximal possible values of space and the time steps  
$\Delta x = 10$ kilometers and $\Delta t=200$ seconds respectively. To obtain the solution 
of the forward problem described above for $96$ hours it is demanded $0.8$ second of 
CPU time. 

\begin{figure}
\begin{center}
\includegraphics[width=10cm]{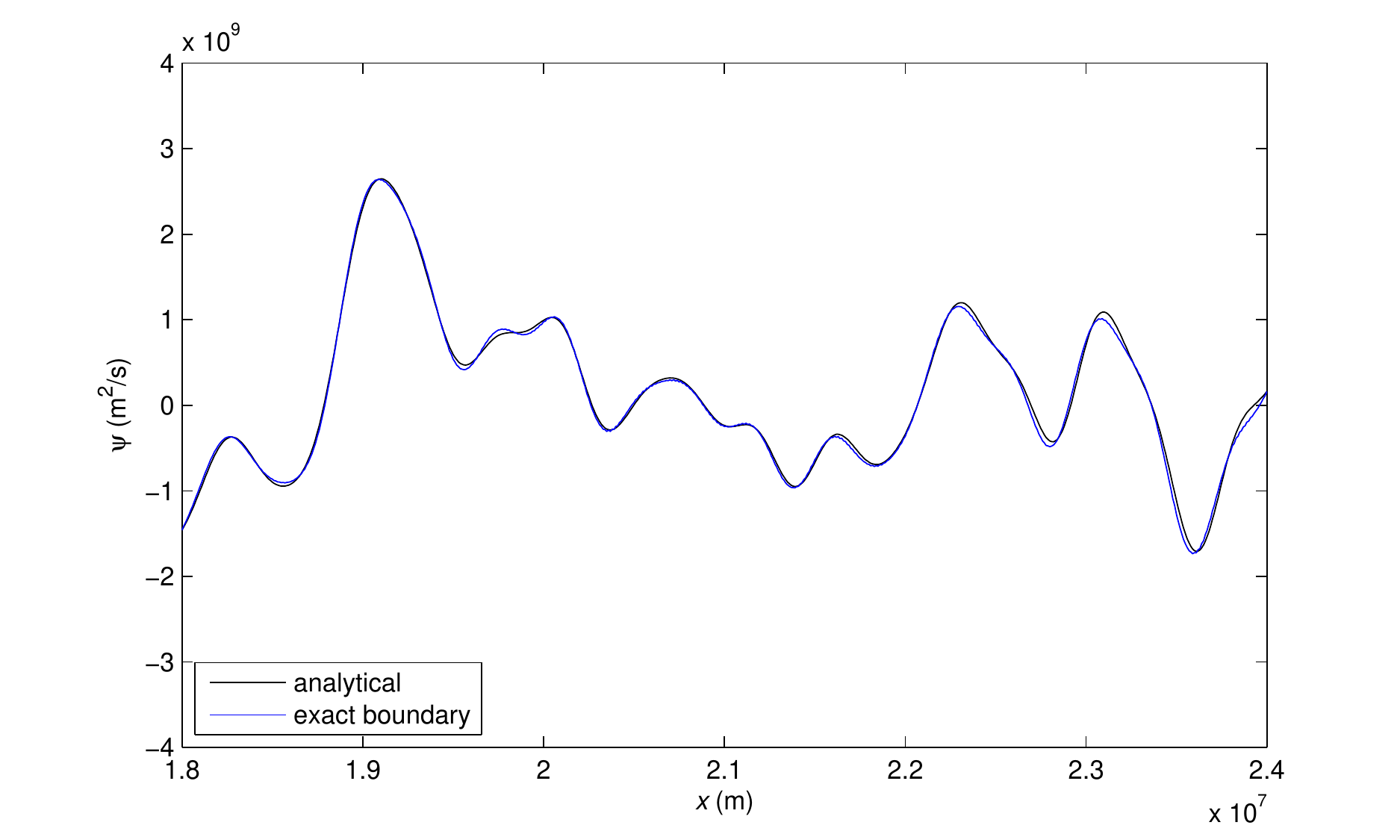}
\caption{\label{fig:fig5} Forward problem solution with exact boundary condition of 
Rossby-Oboukhov equation at $96$ hours. $\Delta x = 10$ km, $\Delta t = 200$ sec. 
CPU time = 0.8 sec.
}
\end{center}
\end{figure}

[Insert figure \ref{fig:fig5} here.]

In the next experiments the perturbed analytical solution on the coarse grid was used 
for formation of the boundary conditions. Figure  \ref{fig:fig5} shows the exact solution 
and the solution with perturbed boundary  conditions of the forward problem.

\begin{figure}
\begin{center}
\includegraphics[width=10cm]{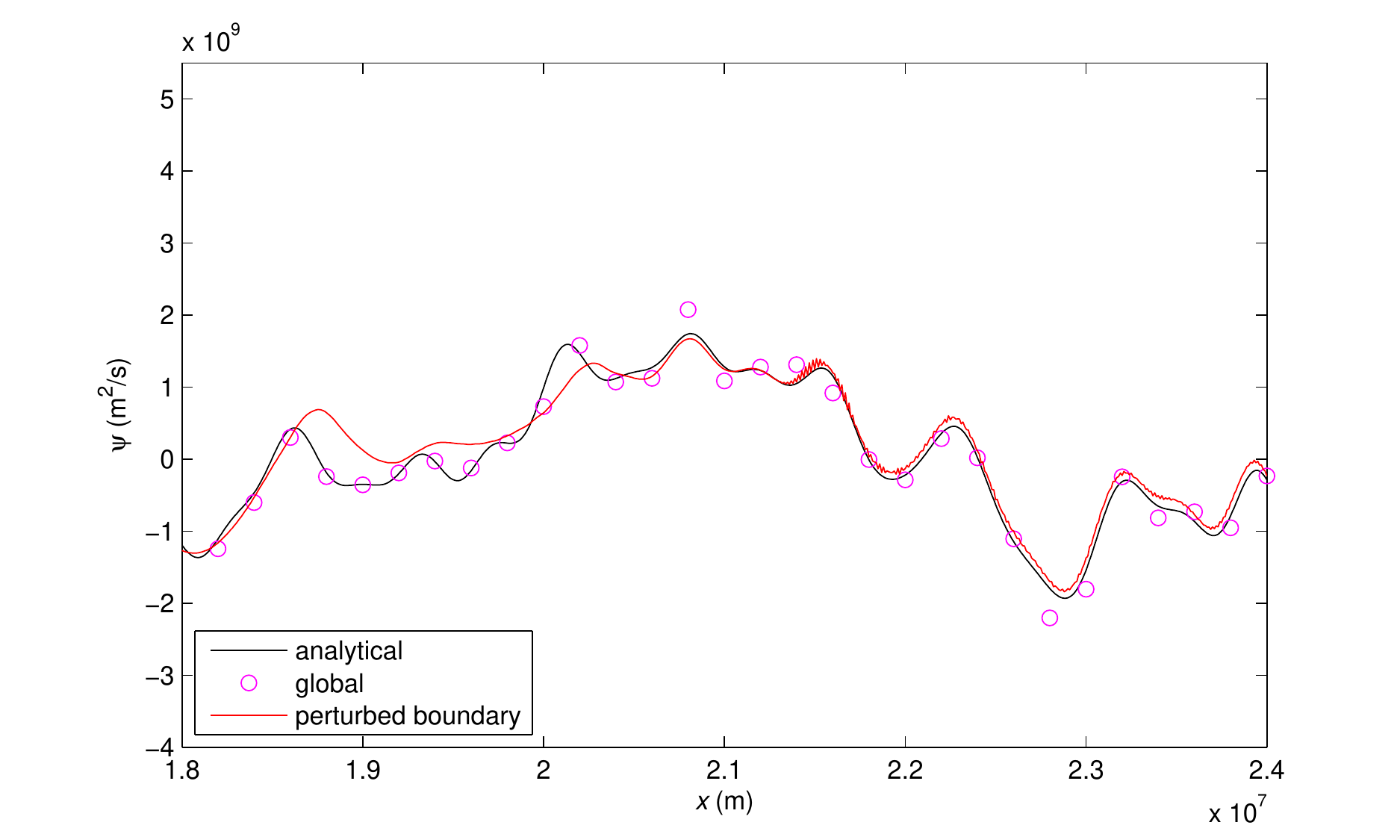}
\includegraphics[width=10cm]{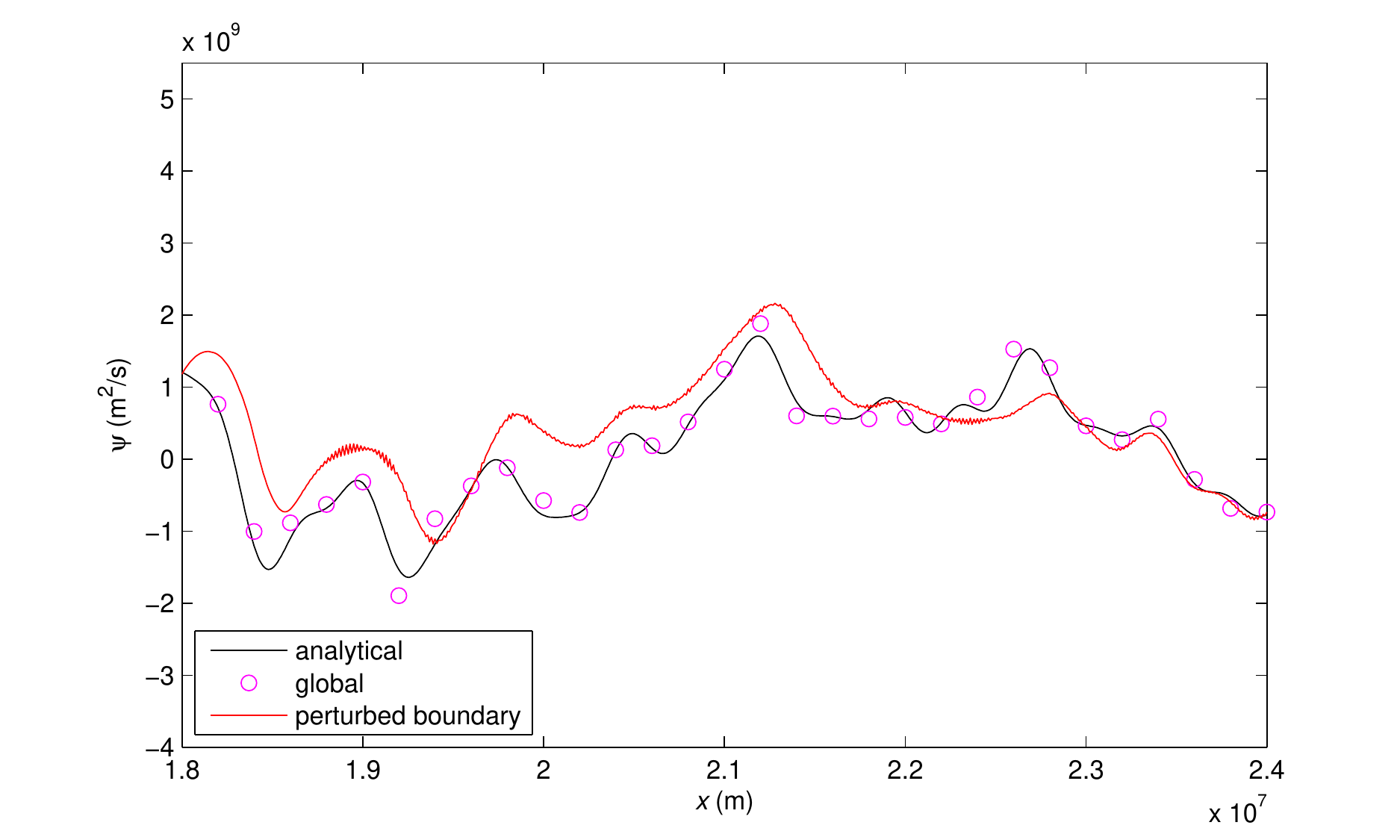}
\includegraphics[width=10cm]{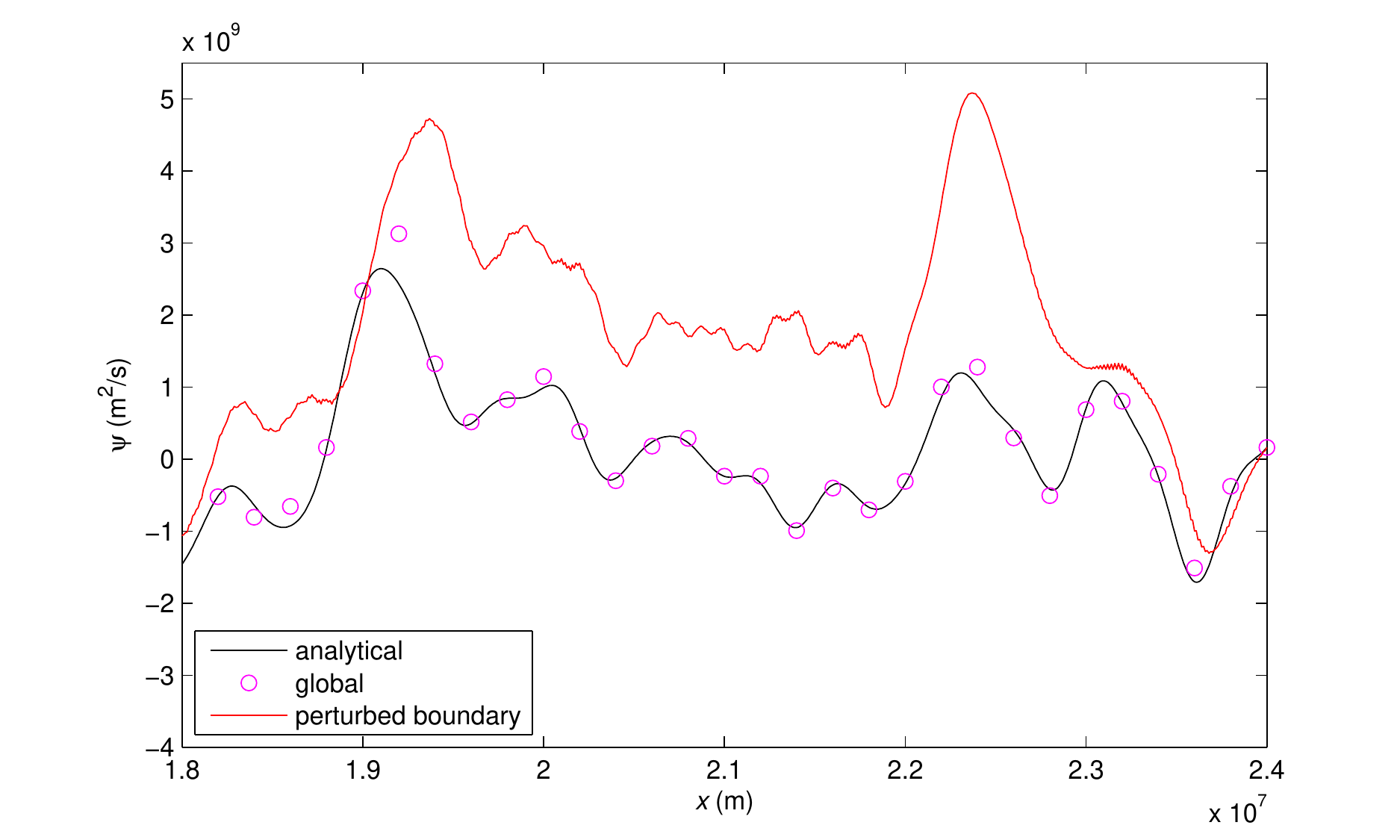}
\caption{\label{fig:fig6} Forward problem solution with 30\% perturbed boundary 
condition of Rossby-Oboukhov equation at $t=24$, $48$, $96$ hours. $\Delta x = 10$ km, 
$\Delta t = 200$ sec. CPU time = 0.8 sec.
}
\end{center}
\end{figure}

[Insert figure \ref{fig:fig6} here.]

One can see that for the forward problem the numerical solution rather quickly diverges 
from the analytical solution due to the errors of the linear interpolation procedure 
in the border points. After  $96$ hours perturbed boundary solution has very faint
resemblance with the analytical one.

In the second series of experiments we applied the optimization method to the local
model. For this calculation we used all available global data (perturbed analytical solution 
on coarse grid) in the inner local domain. Figure \ref{fig:fig6} shows the results of
calculations. It is important to note, that, as in previous experiments, we chose 
the space $\Delta x = 100$ km and the time $\Delta t = 3600$ sec steps to be as large as possible. 

\begin{figure}
\begin{center}
\includegraphics[width=10cm]{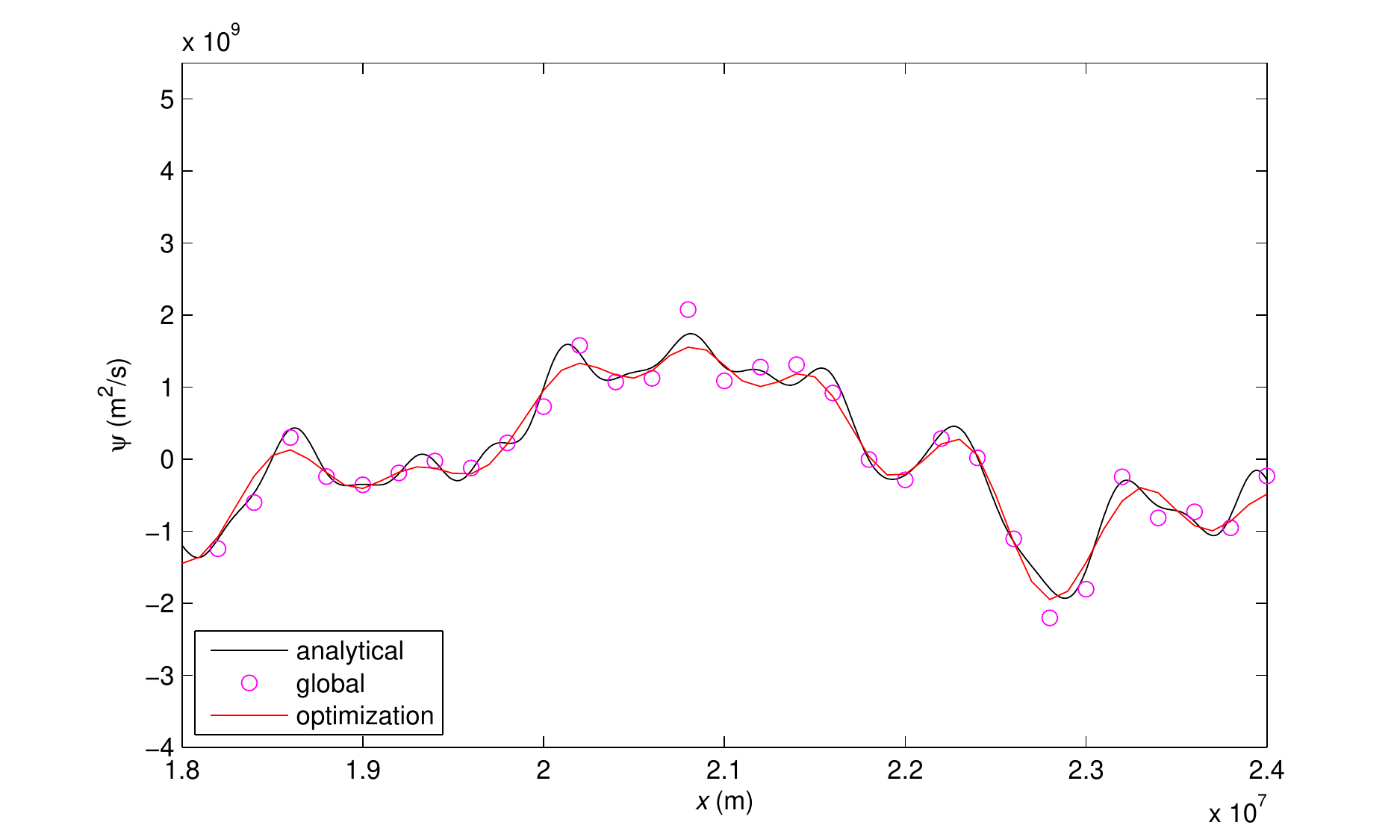}
\includegraphics[width=10cm]{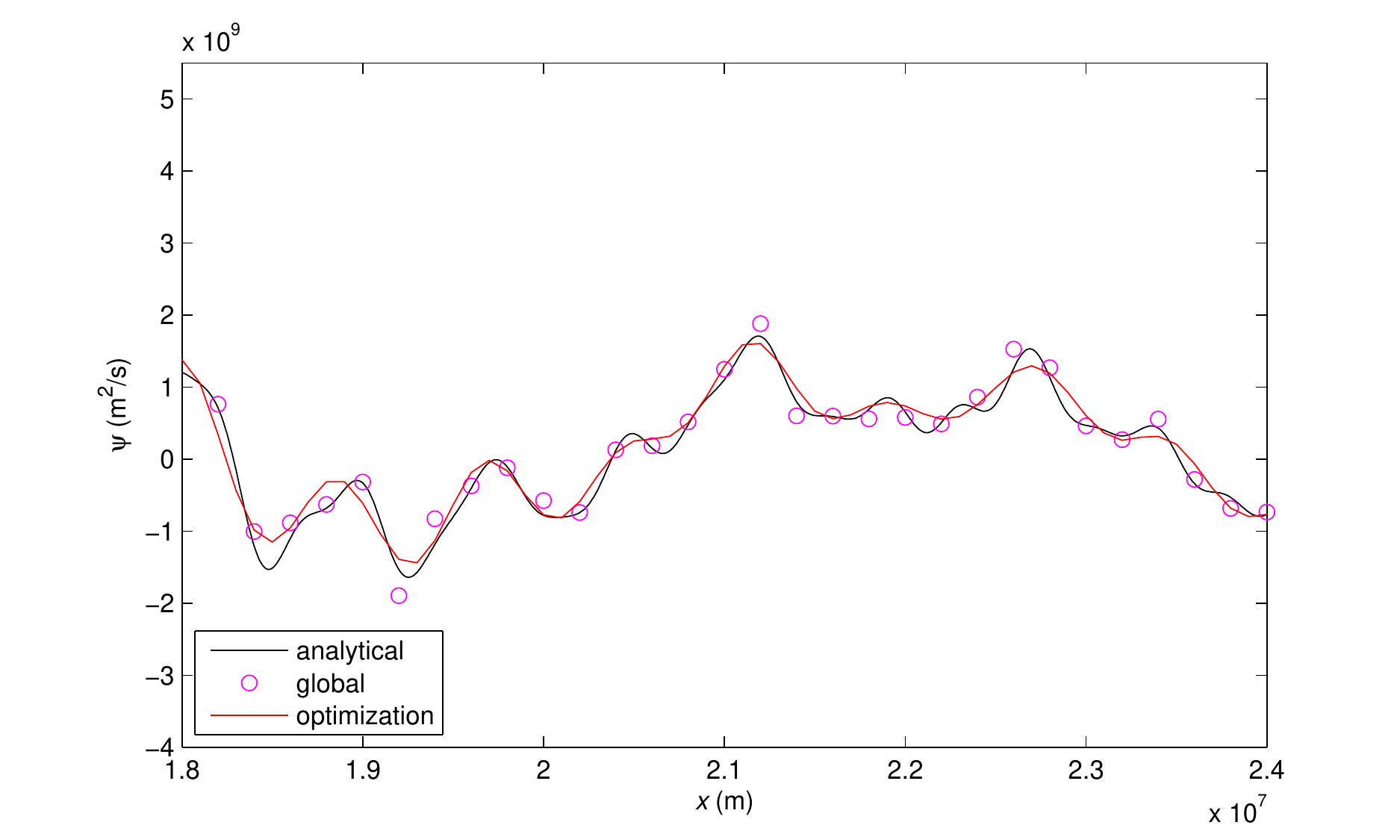}
\includegraphics[width=10cm]{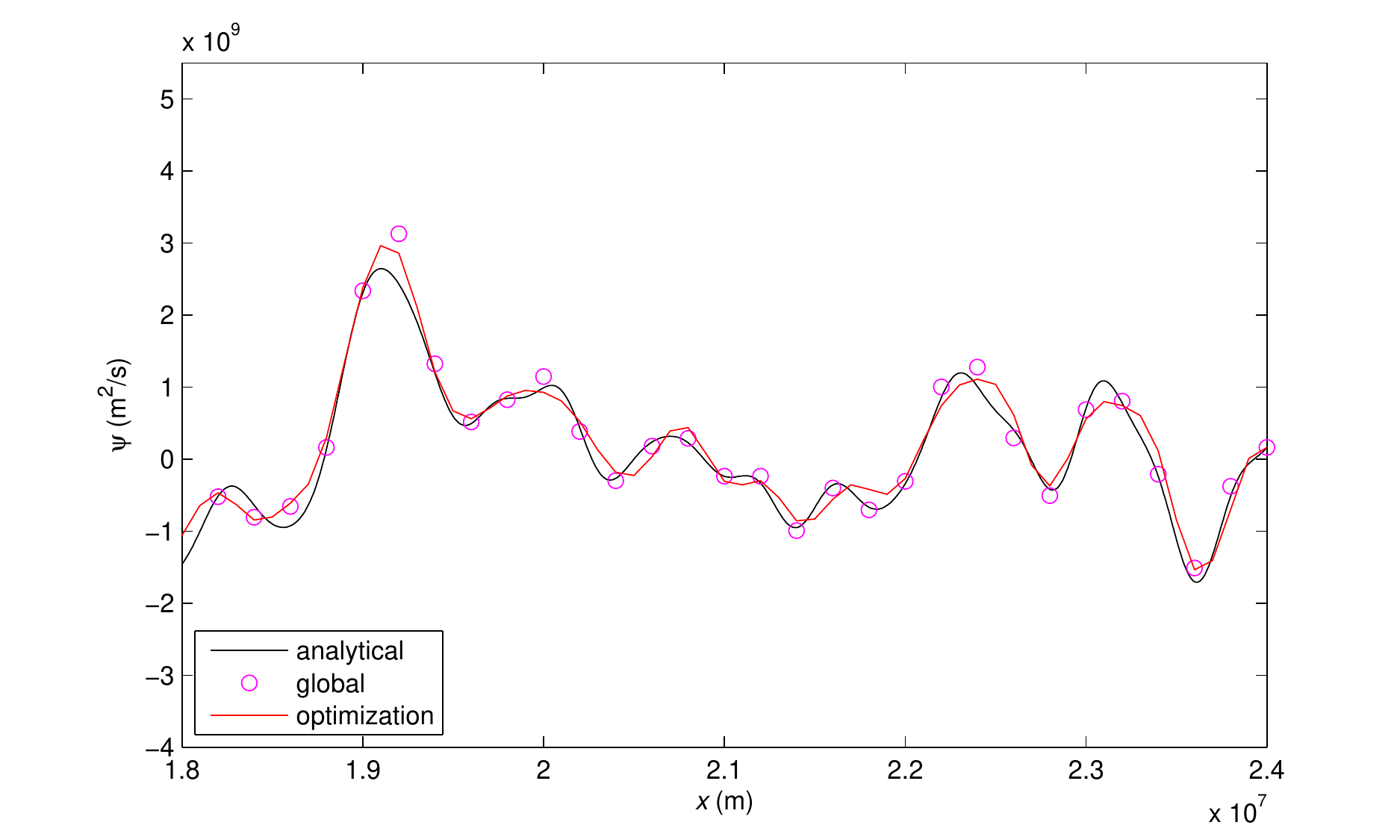}

\caption{\label{fig:fig7} Optimization problem solution of Rossby-Oboukhov equation with 30\% perturbed  
global solution at $t=24$, $48$, $96$ hours. $\Delta x = 100$ km, $\Delta t = 3600$ sec. Average CPU 
time = 0.5 sec.
}
\end{center}
\end{figure}

[Insert figure \ref{fig:fig7} here.]

In the optimization approach one can use larger steps than in the  forward problem, 
because we do not deal with a time-evolution problem, and there is no accumulation of 
numerical errors at each time step. In Figure \ref{fig:fig7}, one can see that for space 
and time steps many times larger than those that were used in the forward problem 
($\Delta x = 100$ kilometers and $\Delta t = 3600$ seconds) we have far better agreement 
with the analytical solution. The decrease of time and/or space step in the optimization 
approach does not appreciably improve the solution, only the smallest details are 
slighly better reproduced. 

\begin{figure}
\begin{center}
\includegraphics[width=10cm]{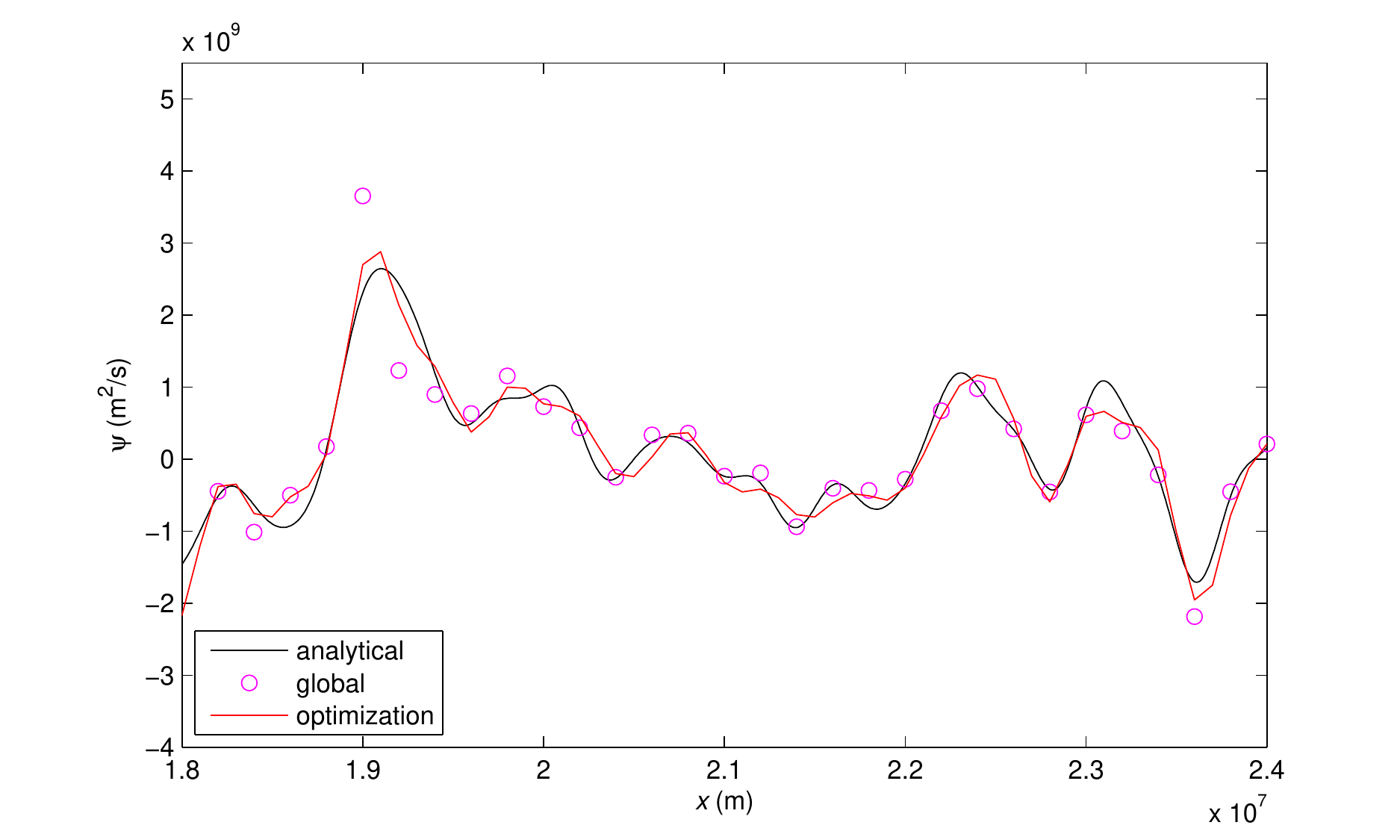}
\caption{\label{fig:fig8} Optimization problem solution of Rossby-Oboukhov equation with 60\% 
perturbed global solution at $96$ hours. $\Delta x = 100$ km, $\Delta t = 3600$ sec. Average CPU 
time = 0.5 sec.
}
\end{center}
\end{figure}

As the local mesh in the optimization approach is rather coarse, the average CPU time
needing for solving the problem is smoller than for forward problem 
($0.5$ seconds) in spite of greater computational complexity. Also we have to pay 
attention on a very weak sensitivity to the global data perturbations. The impact of every 
individual perturbation on the optimization solution is very small. For example, 
increasing errors in randomly perturbed global data up to 60\% does not strongly 
affect the solution as one can see in Figure \ref{fig:fig8}.

[Insert figure \ref{fig:fig8} here.]

These experiments clearly show that the use of additional global data inside the local 
domain, even perturbed, can significantly improve the solution of the model. 
The optimization approach, albeit seeking the solution on entire space-time grid, is not 
very numerically expensive, because the space, and, principally, the time steps can be 
chosen much larger than in the forward problem.

\subsection{Korteweg-de Vries equation}

Now we consider the Korteweg-de Vries equation:
\begin{equation}\label{kdv}
u_t+6uu_x+u_{xxx}=0,
\end{equation}
and find its numerical solution applying both the traditional forward method and 
the optimization  method.

The equation (\ref{kdv}) has the exact solution \citep{korvri95},\citep{grims04}      
\begin{equation}\label{kdv_elliptic}
u=b+a\, \mathrm{cn}^2(\gamma(x-Vt)|m),
\end{equation}
where $\mathrm{cn}(x|m)$ is the Jacobi elliptic function,  $m \, (0 < m < 1) $ is 
the module of elliptic function, $a=2m\gamma^2$ and $V=6b+4(2m-1)\gamma^2$. 
For the case when $m \to 1$ we will have $cn(x|m) \to sech(x)$ and the 
solution (\ref{kdv_elliptic}) will have the form
\begin{equation}\label{kdv_soliton}
u=b+a \, \mathrm{sech}^2(\gamma(x-Vt)),
\end{equation}
with  $V=6b+2a$,and $a=2 \gamma^2$ which describes a one-dimensional soliton .

For the finite-difference discretization we use the following implicit numerical 
scheme \citep{furi99}, which possesses the properties of total energy and 
mass conservation:
\begin{equation*}\label{kdv_impl2}
\begin{split}
&\frac{u_i^{j+1}-u_i^j}{\Delta t} +\frac{1}{2\Delta x}\left((u_{i+1}^{j+1})^2-
(u_{i-1}^{j+1})^2+u_{i+1}^{j+1}u_{i+1}^{j}- \right. \\
&\left. -u_{i-1}^{j+1}u_{i-1}^{j}+(u_{i+1}^{j})^2-(u_{i-1}^{j})^2\right)+
\frac{1}{2\Delta x^3} \left( \frac{u_{i+2}^{j+1}+u_{i+2}^{j}}{2}- \right.\\
&\left. -(u_{i+1}^{j+1}+u_{i+1}^{j}) + (u_{i-1}^{j+1}+u_{i-1}^{j})-
\frac{u_{i-2}^{j+1}+u_{i-2}^{j}}{2}\right)=0.
\end{split}
\end{equation*}
As a numerical scheme is non-linear for obtaining of numerical solution of a forward problem
we use the Newton method.

\begin{figure}
\begin{center}
\includegraphics[width=14cm]{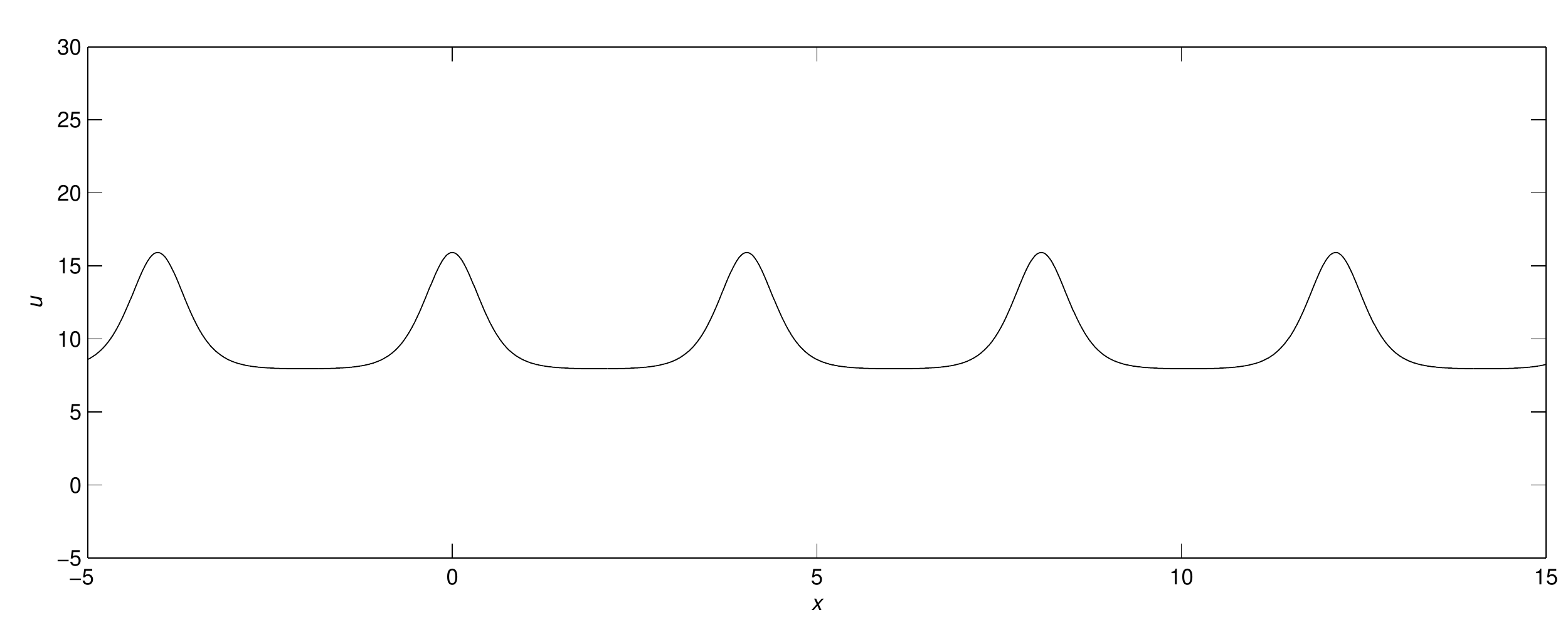}
\caption{\label{fig:fig9}
Analytical solution of KdV equation as the cnoidal wave (\ref{kdv_elliptic}) on 
the domain $[-5,15]$ at $t=0$  with $\gamma=2$, $b=a$ and module $m=0.995$.}
\end{center}
\end{figure}

Following the steps of previos example, we choose as a reference model the analitical 
solution with cofficient $b=a$, $\gamma=2$ and module of elliptic function
$m=0.995$.  The Figure \ref{fig:fig9} represents the solution on domain $[-5,15]$ at 
time $t=0$.

[Insert figure \ref{fig:fig9} here.]

As a local model we consider the equation (\ref{kdv}) defined over closed interval $[0,10]$.
To get a good accordance between a numerical solution of forward problem (with exact initial 
and  boundary conditions) and an anlytical solution on the time interval $0 \le t \le 1$ 
we took as the maximum possible values of space and the time steps $\Delta x=0.02$ and 
$\Delta t = 0.0002$ respectively. The CPU time required to find the solution is 15 seconds.

\begin{figure}
\begin{center}
\includegraphics[width=14cm]{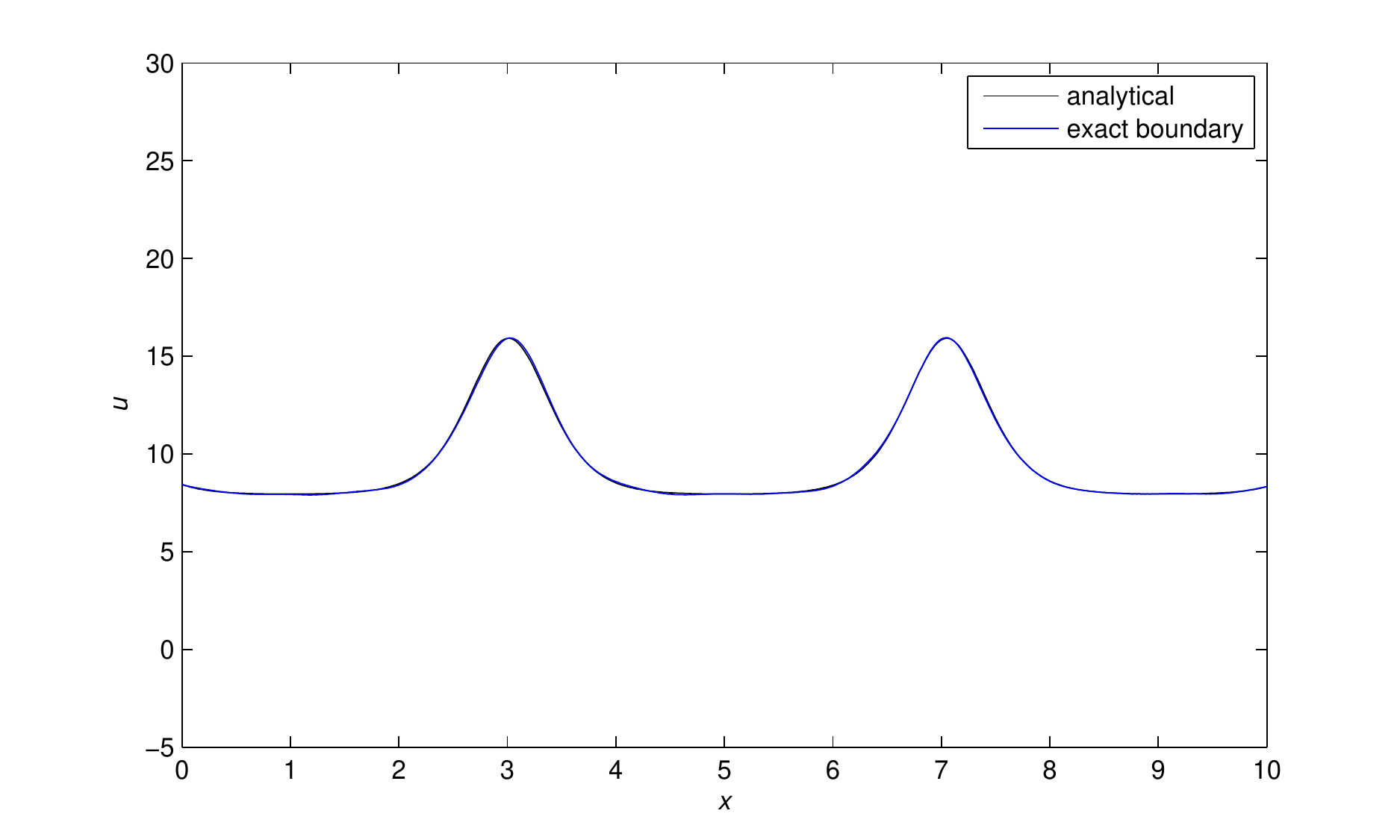}
\caption{\label{fig:fig10}
Forward problem solution with exact boundary condition of KdV equation  at $t=1$.
$\Delta x = 0.02$, $\Delta t = 0.0002$. CPU time = 15 sec.}
\end{center}
\end{figure}

[Insert figure \ref{fig:fig10} here.]

As a global model solution we take the analytical solution, intoduced above as the
reference model, with space step $\Delta x=0.5$ and time step $\Delta t = 0.005$
and perturb its values till $10\%$  from the exact ones.
Using these perturbed global model solution to form  boundary condition for local model 
we will solve initially the forward problem.  

\begin{figure}
\begin{center}
\includegraphics[width=10cm]{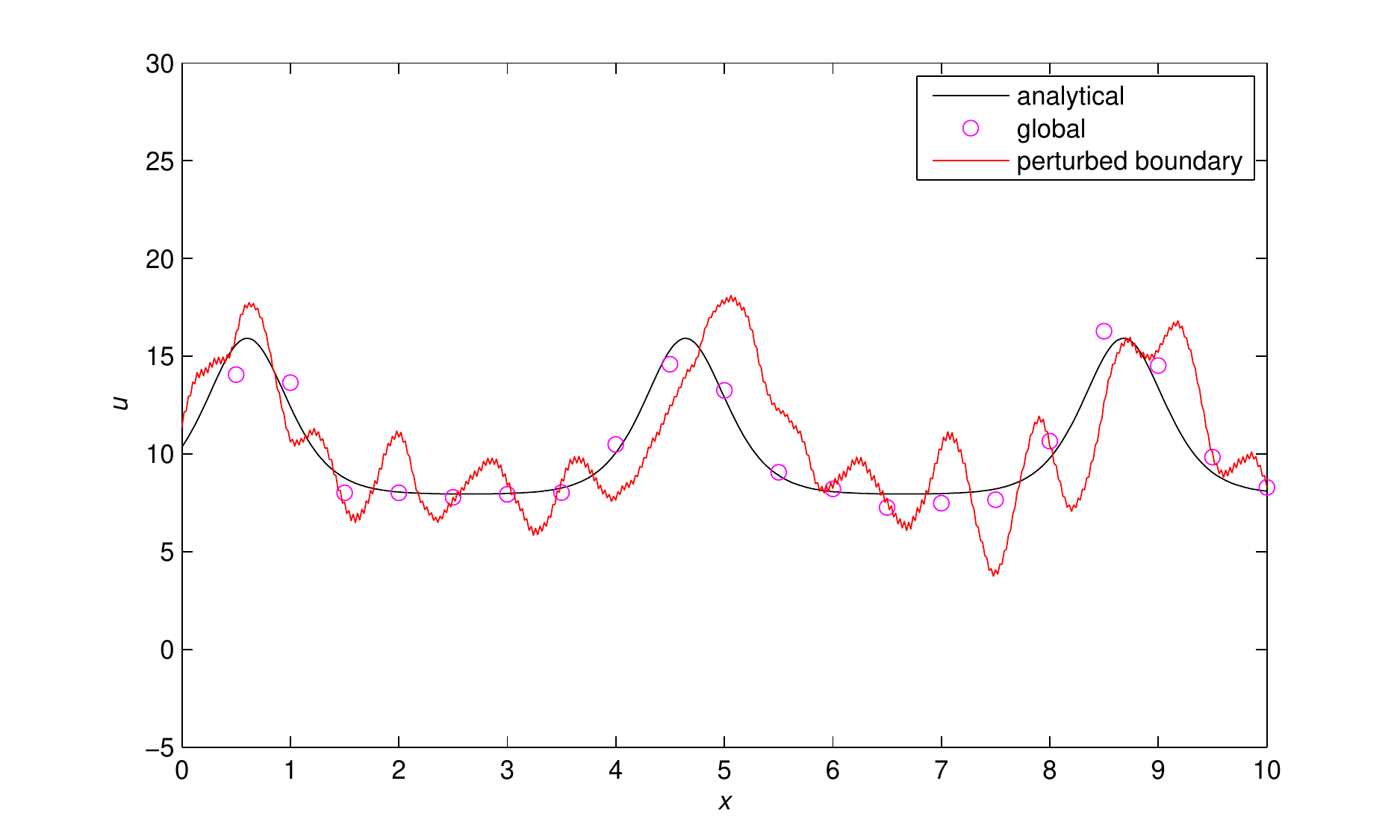}
\includegraphics[width=10cm]{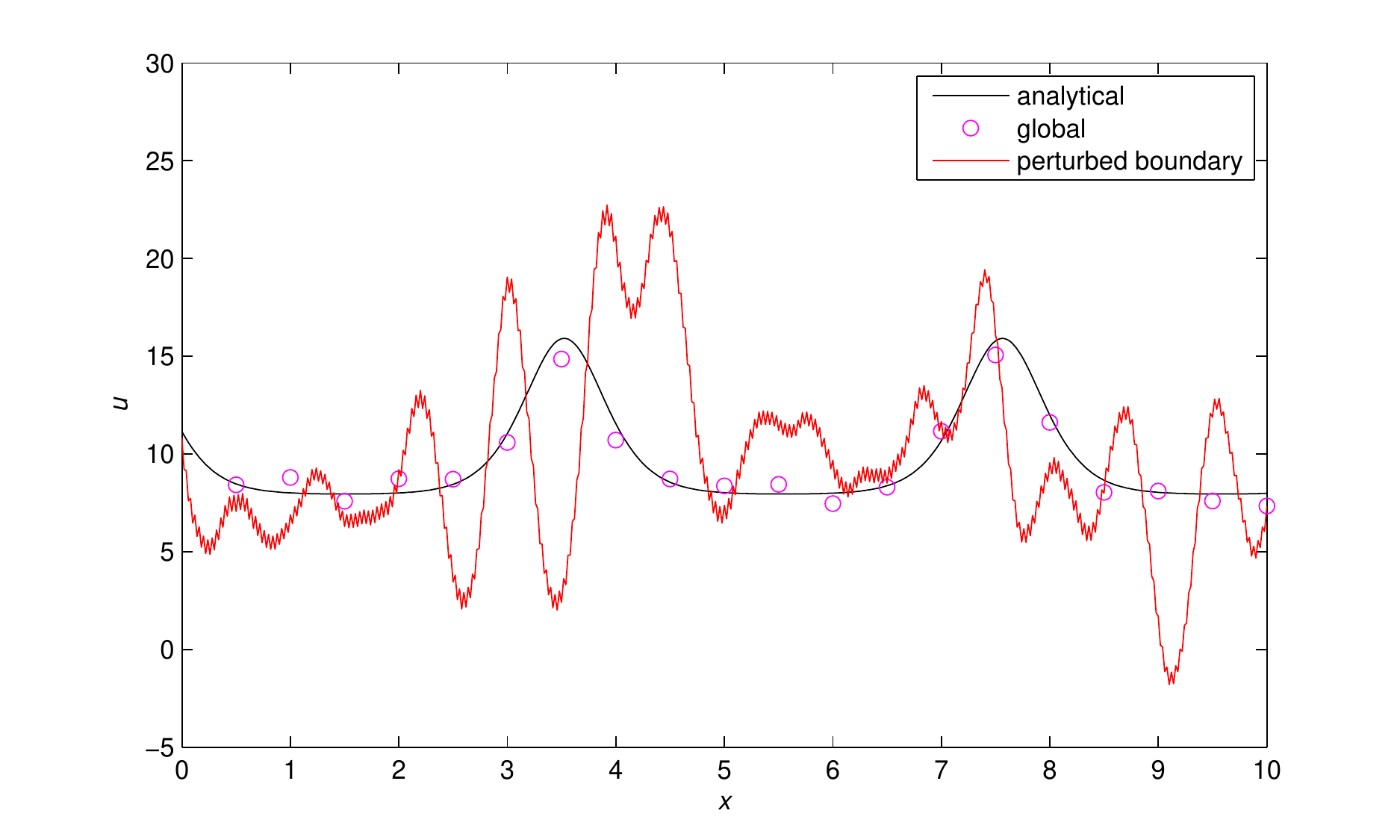}
\includegraphics[width=10cm]{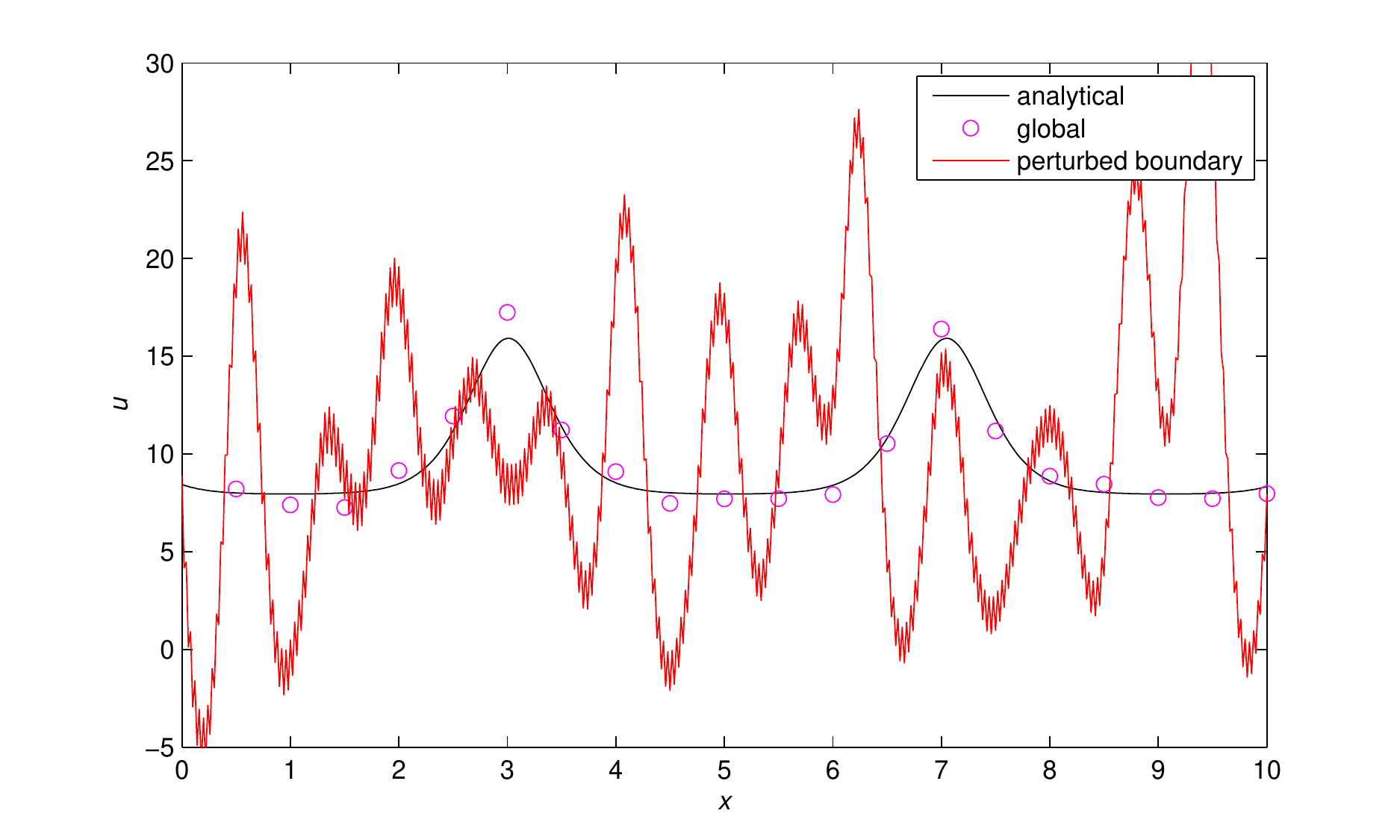}
\caption{\label{fig:fig11} Forward problem solution with 10\% perturbed boundary
condition of KdV equation at time $t=0.2$, $0.5$, $1$. $\Delta x = 0.02$,
$\Delta t = 0.0002$. CPU time = 23 sec.
}
\end{center}
\end{figure}

[Insert figure \ref{fig:fig11} here.]

In Figure \ref{fig:fig11} one can see that the forward problem with small errors in 
the boundary conditions can produce inacceptable numerical solution.

\begin{figure}
\begin{center}
\includegraphics[width=10cm]{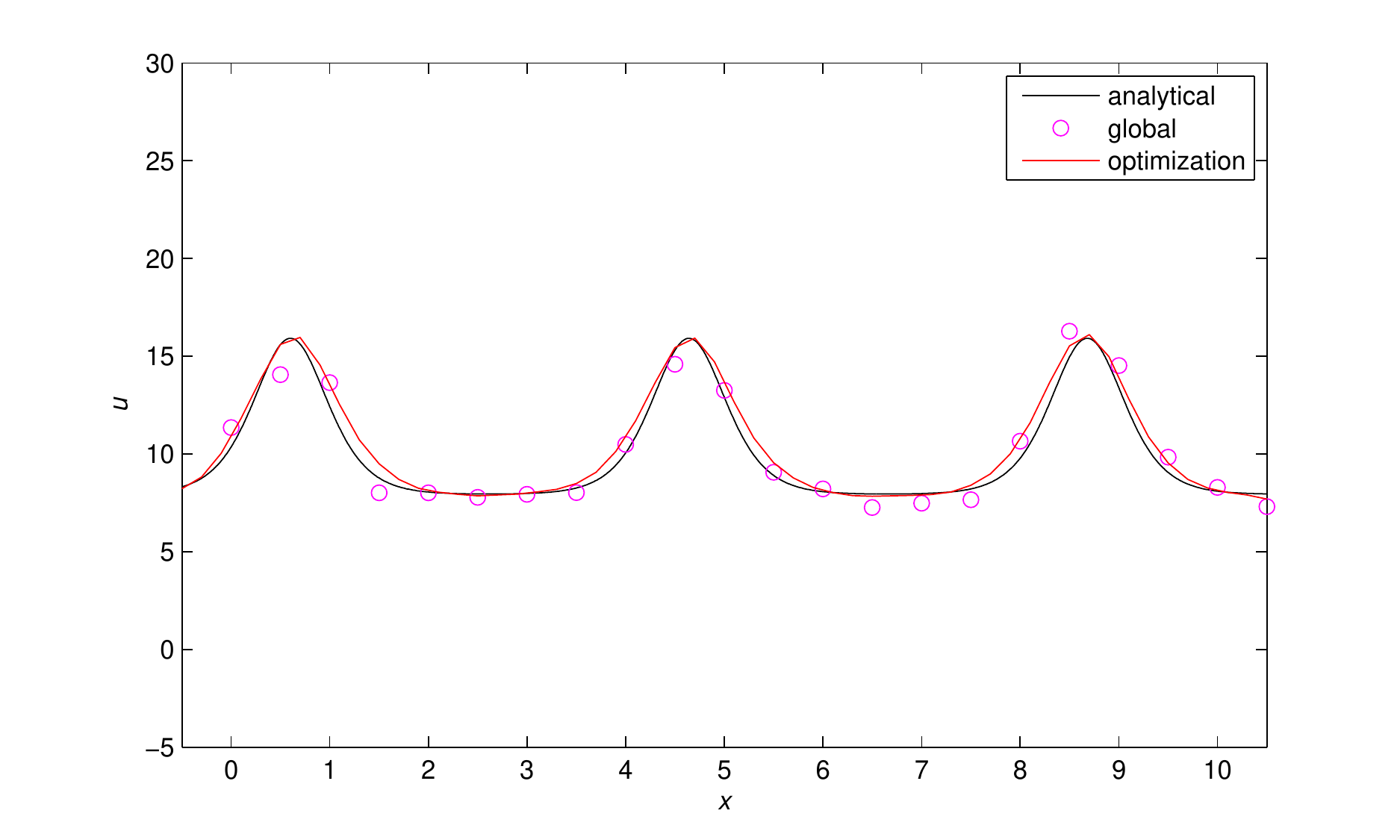}
\includegraphics[width=10cm]{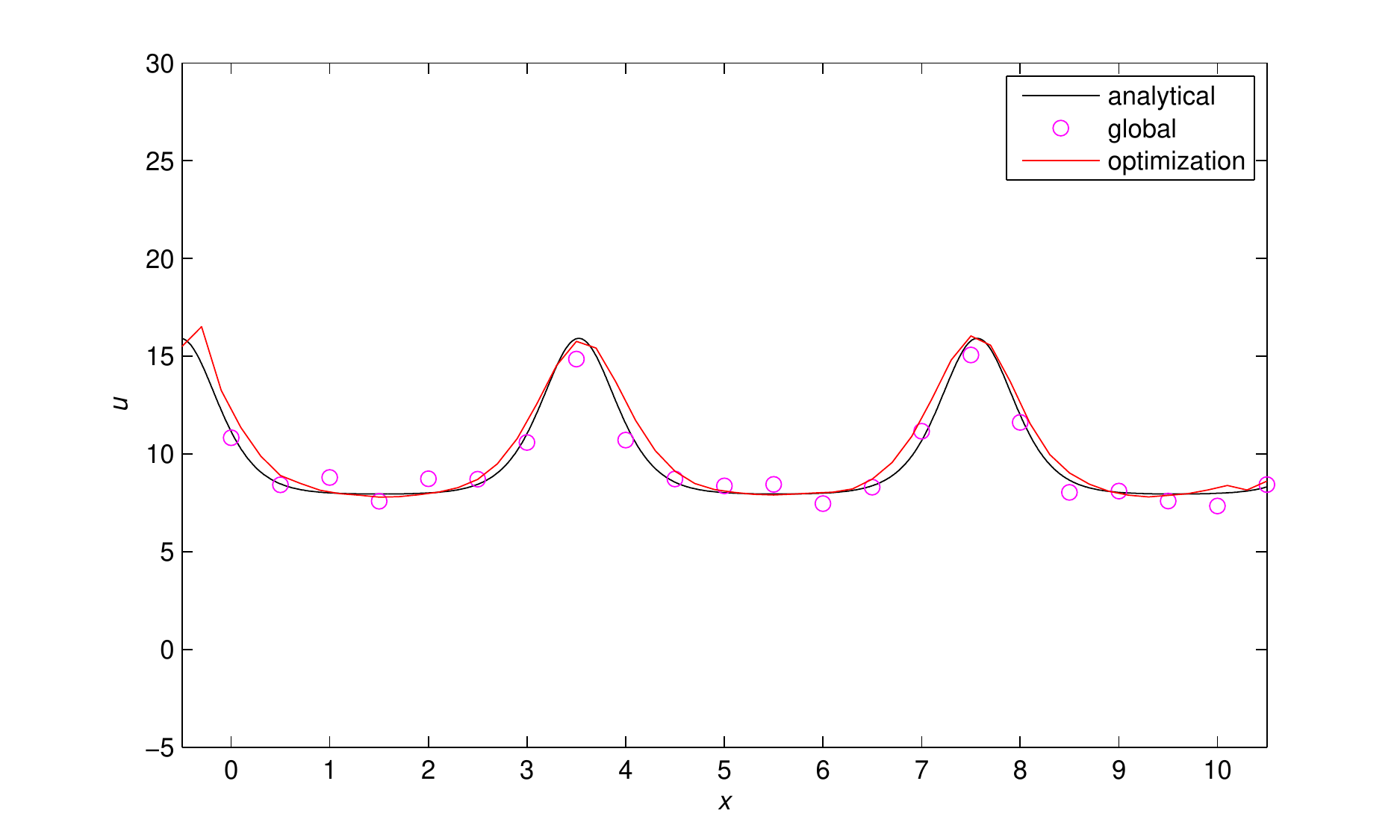}
\includegraphics[width=10cm]{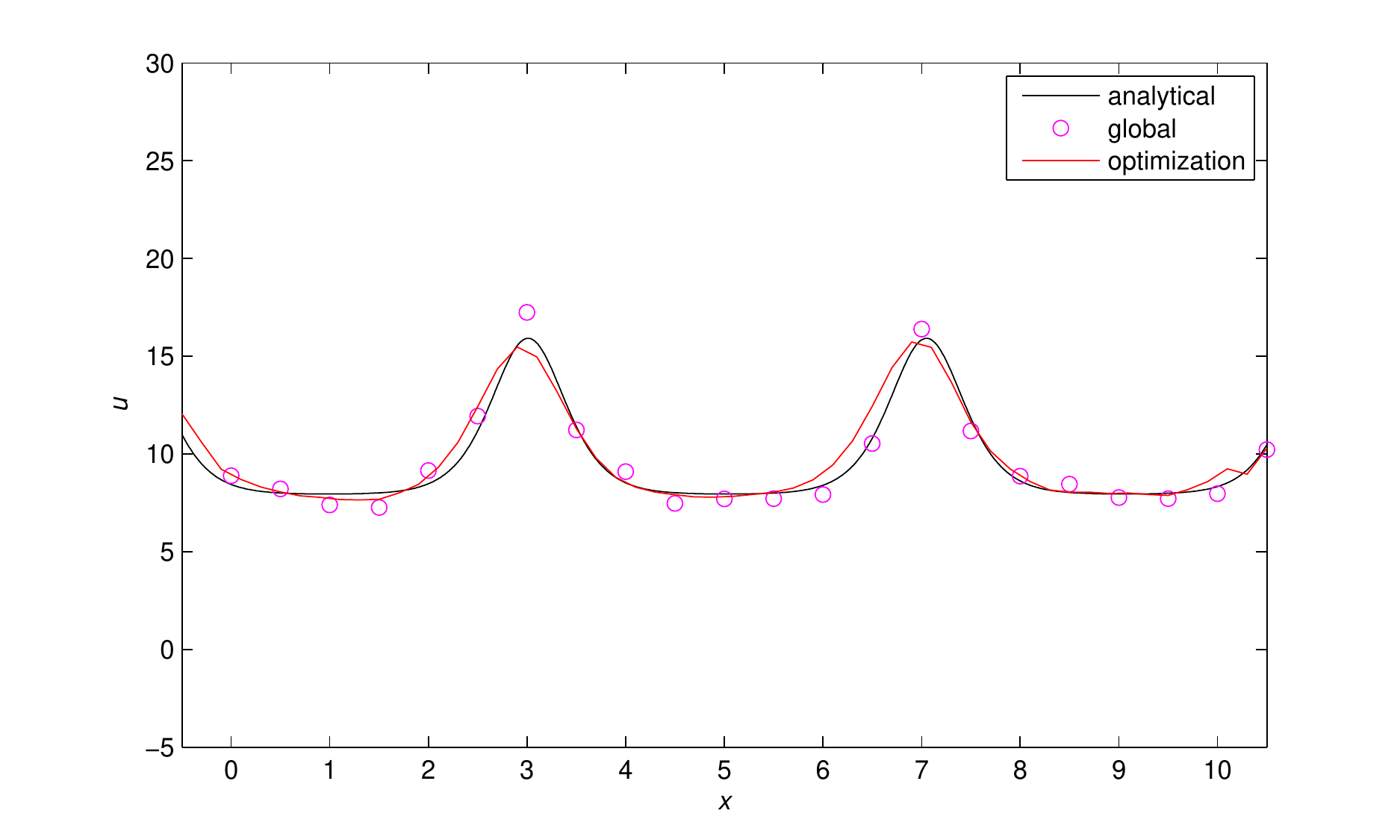}
\caption{\label{fig:fig12} Optimization problem solution of KdV equation with 10\%
perturbed global solution at time $t=0.2$, $0.5$, $1$. $\Delta x = 0.2$, $\Delta t = 0.002$. 
Average CPU time = 21 sec.}
\end{center}
\end{figure}

Now, we apply the optimization method to the local model using additional avaliable 
for inner points information from global model solution. Figure \ref{fig:fig12} shows 
the solution of the optimization problem. As in previous experiment we chose the space
($\Delta x = 0.2$) and the time ($\Delta t = 0.002$) steps as large as possible.

[Insert figure \ref{fig:fig12} here.]

It can be clearly seen the advantage of the optimization approach in this case.
The difference between numerical and exact solution is very small in comparison
with the forward problem calculations.

\section{ Discussions and conclusions}
The results of the numerical experiments presented here indicate that the optimization 
approach can significantly improve the precision of the sought numerical solution of 
the regional model when the boundary values have errors but the information on 
the behaviour of the sought solution in a number of inner points is available.  
Even in the cases in which the solution of Cauchy-Dirichlet problem is very sensitive 
to the errors in the boundary condition the use of optimization approach gives 
the possibility to construct the solution which are close to the analytical solution.
We also made experiments with two-dimentional nonlinear Rossby-Oboukhov equations.
The preliminary results also demonstrate that the use of the optimization approach
significantly improve the numerical solution of boundary problem with errors 
in the boundary values. At the present time we are developing economic algorithm that can be
applied to any great computational problem.

\appendix

\section{\\ \\ \hspace*{-7mm} Formulation of optimization problem}    

We can express our problem as the following non-linear optimization problem 
with equality constraints
\begin{equation}\label{genotim_prob}
\begin{array}{rl}
   \textrm{Minimize} & \frac{1}{2}\|u-V\|^2\\
   \textrm{s.t. }& h(u)=0\textrm{,}
\end{array}
\end{equation}
where $V$ represents the global data on the regional mesh and 
$h(u)= [h_1(u) \, h_2(u) \, \ldots \, h_m(u)]^T$ is a vector 
of the discretizations of regional equations at each point of space-time mesh 
of the regional model.

A usual optimization technique to solve this kind of problem is to apply
Newton's iteration method to the system of nonlinear equations arising from first-order 
necessary conditions, known as Karush-Kuhn-Tucker (KKT) conditions \citep{nocedal99}. 
For instance, the KKT conditions for the problem (\ref{genotim_prob}) are
\begin{equation}\label{kkt_cond}
   \begin{array}{rl}
      (u-V)+h'(u)^{T}\lambda & =0  \\
      h(u) & =0\textrm{,}
   \end{array}
\end{equation}
where
\bd
h'(u)= \left(
   \begin{array}{c}
      \nabla h_1(u)^T \\
      \nabla h_2(u)^T \\
      \vdots \\
      \nabla h_m(u)^T
   \end{array}
   \right)
\ed
is the Jacobian matrix of the constraint function and $\lambda$ represent the vector 
of  Lagrange coefficients. Each step of the Newton iteration  associated with 
system (\ref{kkt_cond}) is defined as follows:
\begin{equation}\label{newton_step}
   \begin{array}{c}
      J(u_k,\lambda_k)
      \left(
         \begin{array}{c}
            \Delta u_k \\
            \Delta \lambda_k
         \end{array}
      \right)= 
      -\left(
         \begin{array}{l}
            u_k-V+h'(u_k)^{T}\lambda_k \\
            h(u_k)
         \end{array}
      \right)\\\\
      \left(
         \begin{array}{c}
            u_{k+1} \\
            \lambda_{k+1}
         \end{array}
      \right)=
      \left(
         \begin{array}{c}
            u_{k} \\
            \lambda_{k}
         \end{array}
      \right)+
      \left(
         \begin{array}{c}
            \Delta u_{k} \\
            \Delta \lambda_{k}
         \end{array}
      \right)
   \end{array}
\end{equation}
where $J$ represents the Jacobian matrix of the system (\ref{kkt_cond})
\begin{equation}\label{jacobian}
   J(u,\lambda)=\left(
   \begin{array}{ccc}
      I+\sum_{i}\lambda_{i}\nabla^{2}h_{i}(u) & & h'(u)^{T} \\\\
      h'(u) & & 0
   \end{array}
   \right)\textrm{,}
\end{equation}
and $\nabla^{2}h_{i}(u)$, $i=1,\ldots,m$, are Hessian matrices of the constraints.

The Jacobian matrix (\ref{jacobian}) is a saddle point matrix and
there are many methods that can be applied with assosiated linear system
\citep{benzi05}. But first, note that the derivative of the discretization 
vector of the constraints is described by a sparse matrix with block-diagonal 
structure, because this derivative is calculated at every point of time-space 
mesh. On the other hand, the first part $I+\sum_{i}\lambda_{i}\nabla^{2}h_{i}(u)$ 
of Jacobian matrix (\ref{jacobian}) of the system (\ref{kkt_cond}) includes 
the calculation of the Hessians of the constraints discretization vector, 
which is a computationally expensive procedure because calculations 
have to be made for every Newton iteration. The resulting matrix is 
generally dense. To accelerate the calculations and reduce the consuming
of the computer memory we use the following Jacobian approximation:

\begin{equation}\label{jacobian_approx}
   B_k=\left(
      \begin{array}{cc}
         I & h'(u_k)^{T}\\
         h'(u_k) & 0
      \end{array}
   \right).
\end{equation}

The use of the approximation $B_k$ for $J(u_k,\lambda_k)$  
strongly simplifies the procedure of finding the solution of the linear 
system (\ref{newton_step}) \citep{benzi05}.

\begin{acknowledgements}
The authors gratefully acknowledge Prof. R. Laprise for constructive discussions with whom
helped to make clear a number of diffcalties of studing problem and Dr. T.A. Tarasova 
for useful comments on the text that significantly improved the manuscript.
\end{acknowledgements}

\end{document}